\documentclass[10pt,prc,aps]{revtex4}
\draft
\usepackage{graphicx}
\usepackage{color}
\usepackage{mathtools}
\begin{document}

\title{The Symmetry Energy:  \\
Current Status of {\it Ab Initio} Predictions {\it vs.} Empirical Constraints }
\author{            
Francesca Sammarruca\footnote{ fsammarr@uidaho.edu }   }                                                        
\affiliation{ Physics Department, University of Idaho, Moscow, ID 83844-0903, U.S.A. 
}
\date{\today} 
\begin{abstract}
Infinite nuclear matter is a suitable laboratory to learn about nuclear forces in many-body systems.  Modern theoretical predictions of neutron-rich matter are particularly timely in view of recent and planned measurements of observables which are sensitive to the equation of state of isospin-asymmetric matter. For these reasons, over the past several years we have taken a broad look at the equation of state of neutron-rich matter and the closely related symmetry energy, which is the focal point of this article. Its density dependence is of paramount importance for a number of nuclear and astrophysical systems, ranging from neutron skins to the structure of neutron stars. We review and discuss {\it ab initio} predictions in relation to recent empirical constraints. We emphasize and demonstrate that free-space $NN$ data pose stringent constraints on the density dependence of the neutron matter equation of state, which essentially determines the slope of the symmetry energy at saturation. 
\end{abstract}
\maketitle 
 
\section{Introduction} 
\label{Intro} 

Although infinite nuclear matter is an idealized system, its equation of state (EoS) is a powerful tool for exploring nuclear interactions in the medium. Asymmetric nuclear matter is characterized by the degree of neutron excess, all the way to pure neutron matter. 
Because neutrons do not form a bound state, the presence of excess neutrons in a nucleus reduces the binding energy, namely, it is a necessary but destabilizing effect which gives rise to the symmetry energy. As a consequence, neutron-rich structures have common features, which explain the formation of a neutron skin (in isospin-asymmetric nuclei) or how neutron stars are supported against gravitational collapse by the outward pressure existing in dense systems with high neutron concentration.  Studies of nuclear interactions in systems with high or extreme neutron to proton ratio are crucial towards understanding of the neutron driplines, the location of which is not well known. The new Facility for Rare Isotope Beams (FRIB), operational since May 2022, is expected to increase the number of known rare isotopes from 3000 to about 6000~\cite{frib}.

Since many years several groups have sought constraints on the density dependence
of the symmetry energy. Intense experimental effort has been and continues to be devoted to this question using various measurements, which are typically analyzed with the help of correlations obtained through different parametrizations 
of  phenomenological models. Popular examples are the Skyrme forces (Refs.~\cite{skyrm1,skyrm2,skyrm3} are only a few of the many review articles on the Skyrme model), and relativistic mean-field models (RMF)~\cite{rmf1,rmf2,rmf3}. Reference~\cite{B_1_2000} and Ref.~\cite{PF_1_2019} are representative examples for applications of Skyrme forces or RMF models, respectively, to the issues confronted in this article.  A cross section of studies that used a variety
 of phenomenological and theoretical models 
or laboratory data to extract constraints on the density dependence of the symmetry energy  
 can be found in Refs.~\cite{SDLD_1_2015, FHPS_1_2010, MCVW_1_2011, MADSCV_1_2017, TLOK17, MADS_1_2018, TRRSWM_1_2018, HY_1_2018, ADS_1_2012, ADSCS_1_2013, VCRW_1_2014, Tsang+09, Tsang+12, LL13, Kort+10, DL_1_2014, Roca+15, Tam+11, Brown13, R_etal_1_2011, R_etal_1_2016}.
 Generally, the extracted constraints vary considerably depending on the methods employed.

During the past two decades, there has been remarkable progress in understanding nuclear forces at a fundamental level, through the concept of effective field theories (EFT)~\cite{Wei90, Wei92}. Meson theoretic or entirely phenomenological models of the nucleon-nucleon ($NN$) interaction, augmented with phenomenological few-nucleon forces, are the typical framework used in the past. At the forefront are state-of-the-art {\it ab initio} predictions, which generally agree on a relatively soft symmetry energy,  in contrast to recent constraints extracted from electroweak scattering experiments~\cite{prexII}, as we will discuss later in the paper.

The density dependence of the symmetry energy is also a major component in the physics of neutron stars, in particular the radius of a medium-mass neutron star. Of course, the structure of a neutron star probes a very large range of densities, from the density
of iron in the outer crust up to several times nuclear matter saturation density in the core, and thus no theory of hadrons can be reliably applied over the entire region. With that in mind,
contemporary {\it ab initio} theories of nuclear and neutron matter at normal densities can be taken as the baseline for any extension or extrapolation method, which will unavoidably
involve some degree of phenomenology. Note, though, that the radius of a 1.4 M$_{\odot}$ neutron star is sensitive to the pressure at normal densities and thus it is a suitable quantity to constrain microscopic theories of the EoS at densities where they are reliable~\cite{Universe}. In fact, the radius of light to medium neutron stars has been found to correlate with the density slope of the symmetry energy at saturation, with correlation coefficient of 0.87 for $ M = 1.0 M_{\odot}$ and 0.75 for  $ M = 1.4 M_{\odot}$~\cite{LR1}.

In this article, we will start with a broad introduction to the symmetry energy and its relevance. To that end, we will briefly review the link between finite nuclei and infinite nuclear matter as it emerges from the liquid drop model. In the process, one defines the energy per nucleon in infinite (isospin-symmetric or asymmetric) matter and the all important symmetry energy. 
 In Sec.~\ref{calc}, we describe our theoretical tools, which include high-quality microscopic nuclear forces derived within the framework of chiral EFT. We also take the opportunity to review the foundations and main aspects of chiral EFT.
 In Sec.~\ref{iso}, after a brief discussion of symmetric nuclear matter (SNM) in Sec.~\ref{snm}, we focus on the neutron matter (NM) EoS and the symmetry energy (Sec.~\ref{pred}). A focal point of this section is a comparison of {\it ab initio} predictions with phenomenological and empirical findings. We include a discussion of the neutron skin, see Sec.~\ref{skin}, and its sensitivity to the density slope of the symmetry energy at saturation. Section~\ref{1s0} reports test calculations designed to underline the sensitivity of the discussed predictions to free-space $NN$ scattering phases.

 In Sec.~\ref{conc}, we reiterate the scope of this article and summarize our conclusions. Also, we wish to reflect on the best way forward to strengthen what should be our most powerful tool --  the link between experiment and {\it ab initio} theory.

\section{The symmetry energy: general aspects}
\label{gen}

We begin with a pedagogical introduction to establish the main concepts and definitions.

The simplest picture of the nucleus goes back to the semi-empirical mass formula (SEMF), also known as the liquid-droplet model. Its ability to capture basic bulk features of nuclei is remarkable in view of its simplicity. The binding energy per nucleon is written in terms of a handful of terms inspired by the dependence of nuclear radii on the cubic root of the mass number A:
\begin{equation}
\label{BoverA}
\frac{B(Z,A)}{A} = a_V - a_{sym}\frac{(A - 2Z)^2}{A^2} - \frac{a_s}{A^{1/3}} - \frac{a_C Z(Z-1)}{A^{4/3}} -\frac{\Delta}{A}  \; ,
\end{equation}
where the last term stands for additional, typically smaller, contributions.
Note that the second term on the RHS depends on the relative neutron-proton asymmetry, or isospin asymmetry, $\frac{N - Z}{A}$, and represents the loss in binding energy experienced by a nucleus due to the destabilizing presence of asymmetry in neutron/proton concentrations. Of course, Eq.~(\ref{BoverA}) is the simplest picture of a nucleus, but can be improved by replacing the number of nucleons $A$ and the number of protons $Z$ with the respective density profiles. 

To that end, we introduce the energy per nucleon, $e(\rho,\alpha)$, in an infinite system of nucleons at density $\rho$ and isospin asymmetry $\alpha = \frac{\rho_n - \rho_p}{\rho}$ -- namely, the EoS of neutron-rich matter -- and expand this quantity with respect to the isospin asymmetry parameter:
\begin{equation}
e(\rho,\alpha) = e(\rho, \alpha=0) + \frac{1}{2} \Big ( \frac{\partial^2 e(\rho,\alpha)}{\partial \alpha^2} \Big )_{(\alpha = 0)}  \alpha^2 + \mathcal{O} (\alpha^{4})  \; .
\label{e_exp}
\end{equation} 
 Neglecting terms of order $\mathcal{O} (\alpha^{4})$, Eq.~(\ref{e_exp}) takes the well-known parabolic form:
\begin{equation}
e(\rho,\alpha) \approx e_0(\rho) + e_{sym}(\rho) \ \alpha^2  \; , 
\label{asym_e}
\end{equation}
where  $e_{sym}$ = 
  $\frac{1}{2} \Big ( \frac{\partial^2 e(\rho,\alpha)}{\partial \alpha^2} \Big )_{\alpha = 0}$ and $e_0(\rho) =  e(\rho, \alpha=0)$, the EoS of isospin-symmetric nuclear matter.

 With the SEMF as a guideline, one can write the main contributions to the total energy of a given nucleus (Z,A) with proton and neutron density profiles $\rho_p(r)$ and $\rho_n (r)$, respectively, as:
\begin{align}
\begin{split}
\label{eee}
E(Z,A)= \int d^3 r \ \rho(r) \ e(\rho,\alpha) + f_0 \int d^3 r  |\nabla \rho|^2  + \frac{e^2}{4 \pi \epsilon_o} (4\pi)^2 \int^{\infty}_{0} dr' \big[ r' \rho_{p}(r') \int^{r'}_{0} dr \ r^2 \rho_{p}(r) \big]  
\end{split} \; ,
\end{align}
where $f_0$ is a constant typically fitted to $\beta$-stable nuclei. Note that the second term on the RHS , dependent on the gradient of the density function, is a finite-size contribution -- the surface term proportional to the coefficient $a_s$ in Eq.~(\ref{BoverA}). The last term on the RHS stands for the Coulomb interaction among protons. The
 link to Eq.~(\ref {BoverA}) is apparent. In particular, the first integral on the RHS of Eq.~(\ref {eee}) comprises the first two terms on the RHS of Eq.~(\ref {BoverA}).

Within the parabolic approximation (Eq.~(\ref {asym_e})), the symmetry energy becomes the difference between the energy per neutron in NM and the energy per nucleon in SNM:
\begin{equation}
e_{sym} (\rho) = e_n (\rho) - e_0 (\rho) \; ,
\label{xxx}
\end{equation}
where $e_n(\rho) =  e(\rho, \alpha=1)$, the energy per neutron in pure neutron matter.

The minimum of $e_0(\rho)$ at a density approximately equal to the average central density of nuclei, $\rho_0$, is a reflection of the saturating nature of the nuclear force. Next, we will
expand the symmetry energy about the saturation point:
\begin{equation}
\label{yyy}
e_{sym} (\rho) \ \approx \ e_{sym} (\rho_{0}) + L \ \frac{\rho -\rho_{0}}{3 \rho_0} + \frac{K_{sym}}{2} \frac{(\rho - \rho_{0})^2}{(3\rho_0)^2}  \; ,
\end{equation}
 which helps identifying several useful parameters. 
 $L$ is referred to as the slope parameter, as it is a measure of the slope of the symmetry energy at saturation:
\begin{equation}
\label{L}
L=3\rho_{0} \Big( \frac{\partial e_{sym}(\rho)}{\partial \rho} \Big)_{\rho_{0}}  \; .
\end{equation}
Furthermore, it is obvious from Eqs. (\ref{xxx}) and (\ref{L}) that $L$ is a measure of the slope of the NM EoS at saturation density, since the SNM EoS has
 a vanishing slope at that point.

 The parameter $K_{sym}$  characterizes the curvature of the symmetry energy at saturation density:
\begin{equation}
K_{sym}=9 \ \rho_{0}^2 \Big( \frac{\partial^2 e_{sym}(\rho)}{\partial \rho^2} \Big)_{\rho_{0}}  \; .
\end{equation}
Note that a similar expansion 
 of the energy per particle in SNM identifies the quantity
\begin{equation}
K_0 = 9 \ \rho_{0}^2 (\frac{\partial^2 e_{0}(\rho)}{\partial \rho^2})_{\rho_{0}}  
\end{equation}
as a measure of the curvature of the EoS in SNM.

Using the standard thermodynamic relation,
\begin{equation}
P(\rho) = \rho^2 \frac{\partial e}{\partial \rho} \; ,
\end{equation}
where $P$ is the pressure and $e$ is the energy per particle, 
 we define the symmetry pressure as:
\begin{equation}
\label{sym_pres}
P_{sym}(\rho) = \rho^2 \frac{\partial (e_{n} - e_0)}{\partial \rho} = P_{NM}(\rho) - P_{SNM}(\rho) \; .
\end{equation}
If the derivative is evaluated at or very near $\rho_0$, the symmetry pressure is essentially the pressure in NM because the pressure in SNM vanishes at saturation. Then:
\begin{equation}
\label{ppp}
P_{NM}(\rho_0) = \Big (\rho^2 \frac{\partial e_n(\rho)}{\partial \rho} \Big )_{\rho_0} \; .
\end{equation}
From  Eq.~(\ref{L}) and Eq.~(\ref{ppp}) it is clear that the slope parameter $L$ is a measure of the pressure in NM around saturation density:
\begin{equation}
\label{L_P}
P_{NM}(\rho_0) = \rho_0 \frac{L}{3}  \; ,
\end{equation}
showing that the pressure in NM is proportional to the slope of the symmetry energy at normal density. The value of $L$ is then a measure of 
the pressure gradient acting on excess neutrons and pushing them outward from the neutron-enriched core of the nucleus to the outer region, thus determining the formation and size of the neutron skin.

\section{Theoretical tools}
\label{calc}

\subsection{Energy per nucleon in infinite matter}
\label{gmat}

To calculate the energy per nucleon in nuclear or neutron matter, we use the nonperturbative particle-particle ladder approximation that provides the leading-order contributions in the hole-line expansion of the energy per particle. 
The next set of diagrams is comprised of the three hole-line contributions, which include
the third-order particle-hole (ph) diagram considered
in Ref.~\cite{Cor+}. The third-order hole-hole (hh) diagram (fourth
order in the hole-line expansion) was found to give a very
small contribution to the energy per particle at normal density
(see Tables II and III of Ref.~\cite{Cor+}). The ph diagram is relatively
larger, bringing in an uncertainty of about 1 MeV on the potential
energy per particle at normal density.

 We compute the
single-particle spectrum for the intermediate-state energies
self-consistently.
The Bethe-Goldstone (or Brueckner) equation for two particles with total center-of-mass momentum $\vec{P}$, initial relative momentum $\vec{q_0}$, and starting energy $\epsilon_0$ in nuclear matter with Fermi momentum $k_F$ is, after angle-averaging of the 
Pauli operator, $Q$:
\begin{equation}
G_{P,e_0,k_F}(q,q_0) = V(q,q_0, k_F) + \int_0^{\infty} dk k^2 \frac{V(q,k,k_F)Q(k,P,k_F)G(k,q_0,k_F)}
{\epsilon_0 - (\epsilon(\vec{P} + \vec{k}) + \epsilon(\vec{P} - \vec{k}) ) + i\delta }  \; .
\label{BG}
\end{equation}
where $V$ is the $NN$ potential augmented with effective three-nuceon forces (3NFs) as density dependent potentials, which will be discussed in Sec.~\ref{III}.
When calculating the single-particle potential, $U$, one must recall that, in order to avoid double-counting, 
the potential $V$ requires a factor of $1/2$ in the density-dependent part at the Hartree-Fock level~\cite{HF}:
\begin{equation}
V = V_{NN} + (1/2)V_{DD} \; ,
\label{HF}
\end{equation} 
where, again, the first term on the right-hand side is the actual $NN$ potential and the second one stands for a density-dependent effective 3NF.

\begin{figure*}[!t] 
\centering
\hspace*{-1.0cm}
\includegraphics[width=10.0cm]{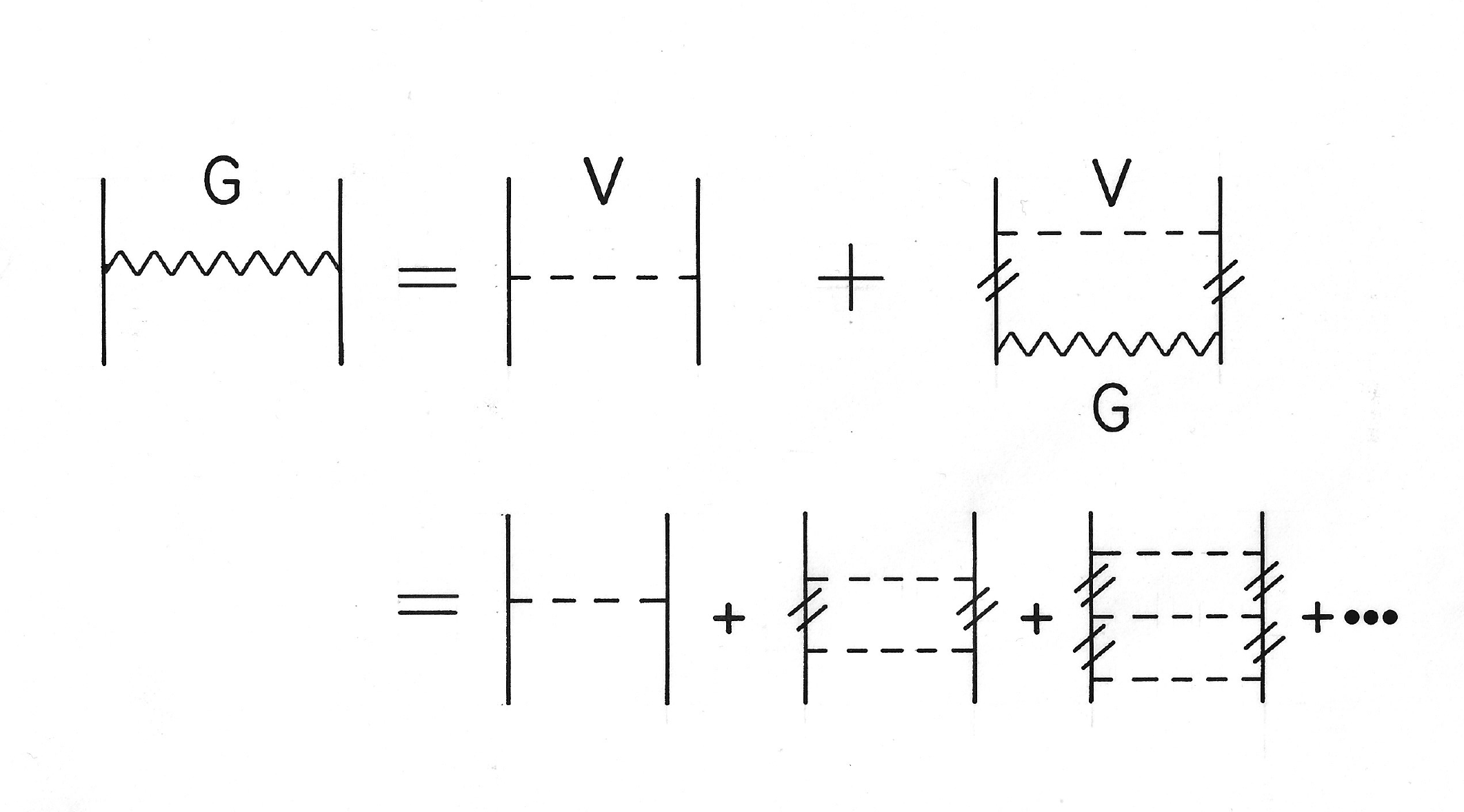}\hspace{0.01in} 
\vspace*{-0.2cm}
\caption{Diagrammatic representation of the Brueckner integral equation. Intermediate nucleon lines with the double slash represent in-medium particle states. 
}
\label{ggg}
\end{figure*}

Using partial wave decomposition, Eq.(\ref{BG}) can be solved for each partial wave using standard matrix inversion techniques. Note that the single-particle energy, $\epsilon$, contains the single-particle potential, $U$,  yet to be determined. Because the G-matrix depends on $U$ and $U$ depends on $G$, an iteration scheme is applied to obtain a self-consistent solution for $U$ and $G$, which are related as:
\begin{equation}
\label{uuu} 
U(k_{1},k_F) = \int  \ \langle \vec{q_{o}} | G_{P,e_0,k_F} | \vec{q_{o}} \rangle d^{3} k_{2}(\vec{q_{o}}, \vec{P})  \; ,
\end{equation}
where $k_{1,2}$ are sincle-particle momenta. Starting from some initial values and a suitable parametrization of the single-particle potential, a first solution is obtained for the G-matrix, which is then used in Eq.(\ref{uuu}). The procedure continues until convergence to the desired accuracy. A diagrammatic representation of the G-matrix equation in the ladder approximation is shown in Fig.~\ref{ggg}.

The energy per nucleon is then evaluated as:
\begin{equation}
\frac{E}{N} = \langle T (k_{1}) \rangle_{(k_{F})} + \langle U (k_{1},k_F) \rangle_{(k_{F})} \; ,
\end{equation}
where the averages of the single-nucleon kinetic ($T(k_{1})$) and potential ($U(k_{1},k_F)$) energies are taken over the Fermi sea. 

 Next, we give a short review of the main principles and advantages of chiral effective field theory (EFT) and describe the input two-nucleon forces (2NFs) and 3NFs which we employ.

\subsection{Chiral Effective Field Theory}
\label{eft}

Chiral EFT~\cite{Wei90,Wei92} allows the development of nuclear interactions as an expansion where theoretical uncertainties can be assessed at each order.  The organizational scheme that controls the expansion is known as
``power counting."  

Chiral EFT maintains consistency with the underlying fundamental theory of strong interactions, quantum chromodynamics (QCD), through the symmetries and
symmetry breaking mechanisms of the low-energy QCD Lagrangian.
The first step towards the development of an EFT is the identification of a ``soft scale'' and a ``hard scale.'' For this purpose, guidance can be found in the hadron spectrum, observing the large separation between the mass of the pion and the mass of the vector meson $\rho$. It is therefore natural to identify the pion mass and the $\rho$ mass (approximately 1 GeV) with the soft and the hard scale, respectively. Moreover, since quarks and gluons are ineffective degrees of freedom in the low-energy regime, pions and nucleons can be taken as the appropriate degrees of freedom of the EFT. 

 The link between QCD and the EFT is established through the symmetries of low-energy QCD. Following the prescriptions of the theory as thoroughly reviewed in Ref.~\cite{ME_1_2011}, we begin with the QCD Lagrangian:
\begin{equation}
\mathcal{L} = \bar{q}(i\gamma^{\mu}\mathcal{D}_{\mu} - \mathcal{M})q - \frac{1}{4} \;  \mathcal{G}_{\mu\nu,a}\mathcal{G}^{\mu\nu}_{a} \; ,
\end{equation}
where $q$ is the quark field, $\mathcal{D}_{\mu}$ the gauge covariant derivative, $\mathcal{M}$ the quark mass matrix, and  $ \mathcal{G}^{\mu\nu}_{a}$ the gluon strength field tensor, with $a$ the color index. Note that quark indices (for spin, color, and flavor) have been suppressed. 

Chiral symmetry is conservation of ``handedness," an exact symmetry for massless particles. Thus, the presence of the quark mass matrix term in the above Lagrangian breaks chiral symmetry -- this is the mechanism of {\it explicit} symmetry breaking. 
The quark mass matrix in flavor space,
\begin{equation}
\mathcal{M} = 
\begin{pmatrix}
m_u & 0 \\
0 & m_d 
\end{pmatrix}
\; ,
\end{equation}
 can be written in terms of the identity matrix and the Pauli spin matrix $ \sigma_3$:
\begin{equation}
\mathcal{M} = \frac{(m_{u} + m_{d})}{2}
\begin{pmatrix}
1 & 0 \\
0 & 1
\end{pmatrix}
+\frac{m_{u}-m_{d}}{2}
\begin{pmatrix}
1 & 0 \\
0 & -1
\end{pmatrix}
\; ,
\label{qqq}
\end{equation}
showing that the first term respects isospin symmetry while the second term vanishes if the masses of the ``up" ($u$) and ``down" ($d$) quarks are equal.  We recall that, for massless quarks, the right- and left-handed components do not mix -- the $SU(2)_R \times SU(2)_L$ symmetry, or chiral symmetry. As noted above, the expression in Eq.(\ref{qqq}) breaks chiral symmetry explicitly as a result of the non-zero quark masses. Thus, the small difference in the quark masses breaks isospin symmetry. However, since the masses of the $u$  and $d$ quarks are very small compared to typical hadronic masses, explicit breaking of chiral symmetry is a small effect. We will remain in the lightest quark sector, $u$ and $d$, as appropriate in a theory of pions and nucleons. QCD with three flavors of light quarks, $u$, $d$, $s$, displays $SU(3) \times SU(3)$ global flavor symmetry in the limit of massless quarks. The masses of the heavier quarks are much larger than the QCD spontaneus chiral symmetry breaking scale and thus cannot be treated as a small perturbation around the symmetry limit.

Next, we need to address the {\it spontaneous} breaking of chiral symmetry, for which there is clear evidence in the hadron spectrum. The spontaneous breaking of a global symmetry is accompanied by the appearance of a massless boson, referred to as a ``Goldstone Boson." The particle which fulfills these requirements is the pion, an isospin triplet pseudoscalar boson.
The pion is light relatively to the other mesons in the hadron spectrum but not massless, which is due to the explicit chiral symmetry breaking from the non-vanishing quark masses.

A quick review of the spontaneous symmetry breaking (SSB) mechanism may be useful to the reader. 
The main point is that the state of a sytem (say, the ground state) does not necessarily have the symmetries of the theory from which the state is derived. In the QCD case, a conserved quantity with negative parity (the axial charge, $Q^A_i$) would lead to the expectation that a hadron of positive parity exists for every hadron of negative parity. Inspection of the hadron spectrum reveals that such symmetry -- the existance of degenerate doublets of opposite parity -- is indeed violated. The vector meson with negative parity, $\rho(770)$, cannot be the ``chiral partner" of the $1^+$ $a_1$ meson, which has a mass of 1230 MeV. We recall, however, that the $\rho$ meson exists in three charge (isospin) states, with masses differing by only a few MeV. Therefore, isospin symmetry ($SU(2)_V$) is observed in the hadron spectrum whereas axial symmetry, $SU(2)_R \times SU(2)_L$, is broken.
The spontaneous breaking of a global symmetry generates massless Goldstone bosons, which must have the quantum numbers of the broken symmetry generators, the pseudoscalar $Q^A_i$. Hence, the Goldstone bosons are identified with the isospin triplet of the pseudoscalar pions.

Having identified pions and nucleons as the appropriate degrees of freedom of the EFT, one can proceed to construct the Lagrangian of the effective theory:
\begin{equation}
\label{lagr}
\mathcal{L}_{eff} = \mathcal{L}_{\pi\pi} + \mathcal{L}_{\pi N} + \mathcal{L}_{NN} + ...    \; ,
\end{equation}
which is then expanded in terms of a natural parameter, identified with the ratio of the ``soft scale'' over the ``hard scale'',  $\frac{p}{\Lambda_{\chi}}$, where $p$ is of the order of the soft scale, whereas $\Lambda_{\chi}$ is the energy scale of chiral symmetry breaking, approximately 1 GeV. See, however, Sec.~\ref{error} below for additional comments on the meaning of the chiral symmetry breaking scale, the breakdown scale, and the ultraviolet cutoff applied in calculations with nucleons.
 The contributions to the effective Lagrangian are arranged according to the power counting scheme, with increasing order resulting in smaller terms. While the expansion itself is, of course, infinite, at each order we are assured that the number of terms is finite and the contributions well defined.

 Each order of the chiral expansion is identified with the maximum power of the expansion parameter, $Q = \frac{p}{\Lambda_{\chi}}$, denoted by $\nu$. Note that $p$ in the chiral perturbation expansion stands for the typical value of the momentum in the system under consideration. The first order is dubbed the Leading Order, or ``LO,'' being equivalent to the power $\nu$ = 0. Terms with $\nu$ = 1 vanish as they would violate conservation of parity. The next power in the expansion ($\nu$ = 2) generates the Next-to-Leading-Order (NLO) terms; $\nu$ = 3 gives the Next-to-Next-to-Leading-Order (N$^2$LO), and so on.
At the first two orders of the chiral expansion, only 2NFs are generated, while 3NFs appear for the first time at N$^2$LO.

 To summarize, symmetries relevant to low-energy QCD  are incorporated in the EFT, in particular chiral symmetry. In other words, chiral EFT is a low-energy realization of the strong interaction having the global symmetries of QCD.
Thus, although the degrees of freedom are pions and nucleons instead of quarks and gluons, there exists a solid connection with the fundamental theory of strong interactions.
Chiral EFT employs a power counting scheme in which the progression of 2NFs and 3NFs is constructed in a systematic and internally consistent manner, thus eliminating the inconsistencies which are unavoidable when adopting meson-theoretic or phenomenological forces.
Furthermore, chiral EFT provides a clear method for controlling the truncation error on an order-by-order basis, as we discuss next.

\subsubsection{Quantifying errors in chiral EFT}
\label{error}

 Crucial to chiral EFT is the ability to determine the truncation error.
 If observable $X$ is known at order $\nu$ and at order $\nu+1$, a reasonable estimate of the truncation error at order $\nu$ can be expressed as the difference between the values of the observable at order $\nu$ and at the next order:
\begin{equation}
\Delta X_{\nu} = |X_{\nu+1} - X_{\nu}| \; ,
\label{del} 
\end{equation} 
which is a measure for what is neglected at order $\nu$.
To estimate the uncertainty at the highest (included) order, we follow the prescription of Ref.~\cite{Epel15}. If $p$ is the typical momentum of the system under consideration, one defines $Q$ as the largest between $\frac{p}{\Lambda_{b}}$ and $\frac{m_{\pi}}{\Lambda_{b}}$, where $\Lambda_{b}$ is the breakdown scale, taken to be about 600 MeV~\cite{Epel15}. Before proceeding, some comments are in place to avoid confusion. In the pion-nucleon sector, it's natural to set the scale to the chiral symmetry breaking scale, $\Lambda_{\chi}$, about
 4$\pi F_{\pi} \approx$ 1 GeV, where $F_{\pi}$ is the pion decay constant, equal to 92 MeV (see Ref.~[21] of Ref.~\cite{Epel15}). However, in the nucleon sector, it is common practice to apply a so-called breakdown scale, $\Lambda_b$, chosen around 600 MeV. This scale is smaller than $\Lambda_{\chi}$ because the non-perturbative resummation, necessary for nucleons, fails for momenta larger than approximately 600 MeV.

 The uncertainty in the value of observable $X$ at N$^3$LO as derived in Ref.~\cite{Epel15} can be understood with the following arguments. If N$^3$LO ($\nu$ = 4) is the highest included order, the expression 
\begin{equation}
\Delta X_4 = |X_{4} - X_3| Q =( \Delta X_3) Q \; ,
\label{del43} 
\end{equation} 
is a reasonable estimate for $\Delta X_4$ in absence of the value $X_5$, because $Q$ to the power of 1 takes the error up by one order, the desired 4th order. To avoid accidental underestimations, a more robust prescription is to proceed in the same way for all the lower orders ($\nu$ = 0, 2, 3) and define, at N$^3$LO:
\begin{displaymath}
\Delta X = \max \{Q^5|X_{LO}|, Q^3|X_{LO} - X_{NLO}|,Q^2|X_{NLO} - X_{N^2LO}|, 
\end{displaymath}
\begin{equation}
Q|X_{N^2LO} - X_{N^3LO}| \} \; ,
\label{err}
\end{equation} 
where $p$ could be identified with the Fermi momentum at the density under consideration.

 Cutoff variations have sometimes been used to estimate contributions beyond truncation. However, they do not allow to estimate the impact of neglected long-range contributions. Also, due to the intrinsic limitations of the EFT, a meaningful cutoff range is hard to estimate precisely, and often very limited. The method of Eq.~(\ref{err}) allows to determine truncation errors from predictions at all lower orders, without the need to use cutoff variations.

\subsection{The two-nucleon force}  
\label{II}

The 2NFs we employ are part of a family of potentials from leading order (LO) to fifth order (N$^4$LO)~\cite{EMN17}. The long-range part of the interaction is fixed by the $\pi N$ low-energy constants (LECs) as determined in the very accurate analysis of Ref.~\cite{Hofe+,Hofe++}. Those carry very small uncertainties, as seen from Table II of Ref.~\cite{EMN17}, rendering variations in the $\pi N$ LECs unnecessary when estimating the overall uncertainty in applications of the potential.
Furthermore, at the fifth order, the $NN$ data below pion production threshold are reproduced with excellent precision ($\chi ^2$/datum = 1.15). 

Our $G$-matrix calculation requires iteration of the potential in the non-perturbative Lippmann-Schwinger equation and thus high-momentum components must be suppressed. This is accomplished through the application of a regulator function for which the non-local form is chosen:
\begin{equation}
f(p',p) = \exp[-(p'/\Lambda)^{2n} - (p/\Lambda)^{2n}] \,,
\label{reg}
\end{equation}
where $p' \equiv |{\vec p}\,'|$ and $p \equiv |\vec p \, |$ denote the final and initial nucleon momenta in the two-nucleon center-of-mass system, respectively. $n$ is the exponent of the regulator and is given in Table~\ref{lecs_I}. We use 
 $\Lambda$ = 450 MeV throughout this work, which is reasonable, since it is smaller than the breakdown scale, $\Lambda_b \approx$ 600 MeV.
The potentials we use are relatively soft as confirmed by the 
Weinberg eigenvalue analysis of Ref.~\cite{Hop17} and the perturbative calculations of infinite matter of  Ref.~\cite{DHS19}. In NM we use the neutron-neutron versions of these potentials.

\begin{table*}[t]
\caption{Values of the LECs used in our calculations. $n$ is the exponent of the regulator, Eq.~(\ref{reg}).
The LECs $c_{1,3,4}$ are given in units of GeV$^{-1}$, and $C_S$ and $C_T$ are in units of GeV$^{-2}$. The integer $n$ refers to the exponent of the regulator function, Eq.~(\ref{reg}), with $\Lambda$=450 MeV.} 
\label{lecs_I}
\begin{tabular*}{\textwidth}{@{\extracolsep{\fill}}ccccccc}
\hline
\hline
 Order & $n$ & $c_1$ & $c_3$ & $c_4$   & $C_S$  & $C_T$ \\
\hline    
\hline
N$^2$LO  & 2& --0.74 & --3.61 & 2.44    &         &           \\
N$^3$LO  & 2& --1.07 & --5.32 & 3.56      & --118.13 & --0.25                                \\ 
\hline
\hline
\end{tabular*}
\end{table*}

\subsection{The three-nucleon force} 
\label{III}

Three-nucleon forces first appear at the third order (N$^2$LO) of the $\Delta$-less EFT.  At this order, the 3NF consists of three contributions~\cite{Epe02}: the long-range two-pion-exchange (2PE) graph, the medium-range one-pion-exchange (1PE) diagram, and a short-range contact term. 

In infinite matter, it is possible to construct approximate expressions for the 3NF as density-dependent effective $NN$ interactions~\cite{holt09,holt10} which are represented in  terms of the well-known non-relativistic two-nucleon force operators and, therefore, can be incorporated in the usual $NN$ partial wave formalism and the particle-particle ladder approximation we use for computing the EoS.

\begin{figure*}[!t] 
\centering
\hspace*{-3cm}
\includegraphics[width=7.7cm]{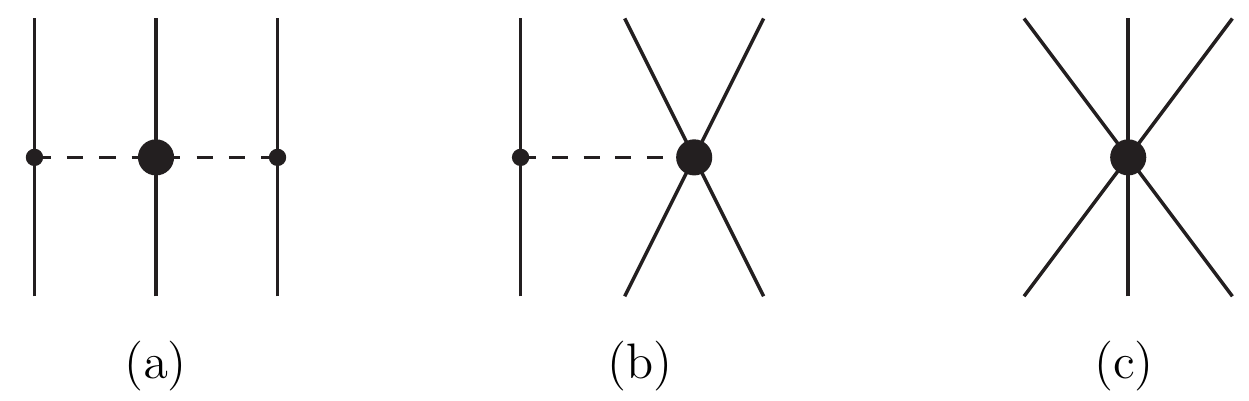}\hspace{0.01in} 
\vspace*{-0.2cm}
 \caption{ Diagrams of the leading 3NF: (a) the long-range 2PE, depending on the LECs $c_{1,3,4}$; (b) the medium-range 1PE, depending on the LEC $c_{D}$; and (c) the short-range contact, depending on the LEC $c_E$.
}
\label{3nfn2lo}
\end{figure*}

 The effective density-dependent two-nucleon interactions at N$^2$LO consist of six one-loop topologies. Three of them are generated from the 2PE graph of the chiral 3NF and depend on the LECs $c_{1,3,4}$, which are already present in the 2PE part of the $NN$ interaction. Two one-loop diagrams are generated from the 1PE diagram, and depend on the low-energy constant $c_D$. The one-loop diagram that involves the 3NF contact diagram depends on the LEC $c_E$. The corresponding diagrams are displayed in Fig.~\ref{3nfn2lo}.
The contributions depending on the LECs $c_D$ and $c_E$ are absent in neutron matter~\cite{HS10}.

\begin{figure*}[!t] 
\centering
\hspace*{-1cm}
\includegraphics[width=12.0cm]{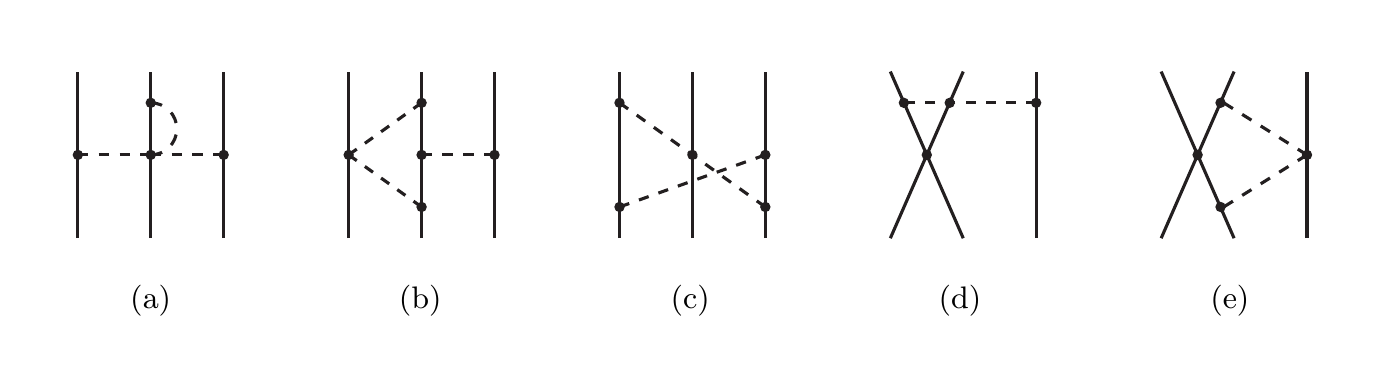}\hspace{0.01in} 
\vspace*{-0.5cm}
 \caption{Some diagrams of the subleading 3NF, each being representative of a particular topology: (a) 2PE; (b) 2P1PE; (c) ring; (d) 1P-contact; (e): 2P-contact. Note that the 1P-contact topology makes a vanishing contribution.
}
\label{3nfn3lo}
\end{figure*}

 The 3NF at N$^3$LO was derived in Refs~\cite{Ber08,Ber11}. The long-range part of 
 the subleading chiral 3NF consists of: the 2PE topology, which is the longest-range component of the subleading 3NF (Fig.\ref{3nfn3lo}(a)), the two-pion-one-pion exchange (2P1PE) topology (Fig.\ref{3nfn3lo}(b)), and the ring topology (Fig.\ref{3nfn3lo}(c)), representing a circulating pion which is absorbed and reemitted from each of the three nucleons. 
The in-medium $NN$ potentials corresponding to these long-range subleading 3NFs are given in Ref.~\cite{Kais19} for SNM and in Ref.~\cite{Kais20} for NM. The short-range subleading 3NF consists of: the one-pion-exchange-contact (1P-contact) topology (Fig.\ref{3nfn3lo}(d)), which gives no net contribution, the two-pion-exchange-contact (2P-contact) topology (Fig.\ref{3nfn3lo}(e)), and relativistic corrections, which depend on the $C_S$ and the $C_T$ LECs of the 2NF and are proportional to $1/M$, where $M$ is the nucleon mass. We include those contributions as well and find them to be very small -- in the order of a fraction of 1 MeV in the energy per nucleon.
The in-medium $NN$ potentials corresponding to the short-range subleading 3NFs are given in Ref.~\cite{Kais18} for SNM and in Ref.~\cite{Treur} for NM.

The LECs we use, which appear already in the 2NF, are displayed in Table~\ref{lecs_I}. We recall the $C_S$ and $C_T$ are determined through the fit of the contact terms in the potential~\cite{EMN17}. The $c_D$ and $c_E$ LECs vanish in NM. Their values for SNM are taken from Ref.~\cite{Suppl}.

 Note that, when the subleading 3NFs are included, the $c_1$ and $c_3$ LECs are replaced by -1.20 GeV$^{-1}$ and -4.43 GeV$^{-1}$, respectively. This is because most of the subleading two-pion-exchange 3NF can be accounted for 
by a shift of the LECs in the leading 3NF equal to -0.13 GeV$^{-1}$ (for $c_1$) and  0.89 GeV$^{-1}$ (for $c_3$)~\cite{Ber08}, plus additional contributions resulting from Eq.~(1) of Ref.~\cite{Kais20}.

\section{Ab initio predictions in infinite matter } 
\label{iso}
We perform  order-by-order calculations, including all subleading 3NFs up to N$^3$LO and with uncertainty quantification as from Eq.(\ref{err}). Thus, we are able to draw reliable conclusions about the convergence pattern of the chiral perturbation series up to fourth order. 

\subsection{Symmetric nuclear matter}
\label{snm}

Although this article is mostly about isospin-asymmetric systems, for completeness we include a short section about SNM, an important laboratory for testing many-body theories. Recently, challenges with simultaneous description of masses and radii of medium-mass nuclei has brought SNM saturation properties to the forefront of contemporary ab initio nuclear structure~\cite{fake}.

\begin{figure*}[!t] 
\centering
\hspace*{-1.0cm}
\includegraphics[width=7.8cm]{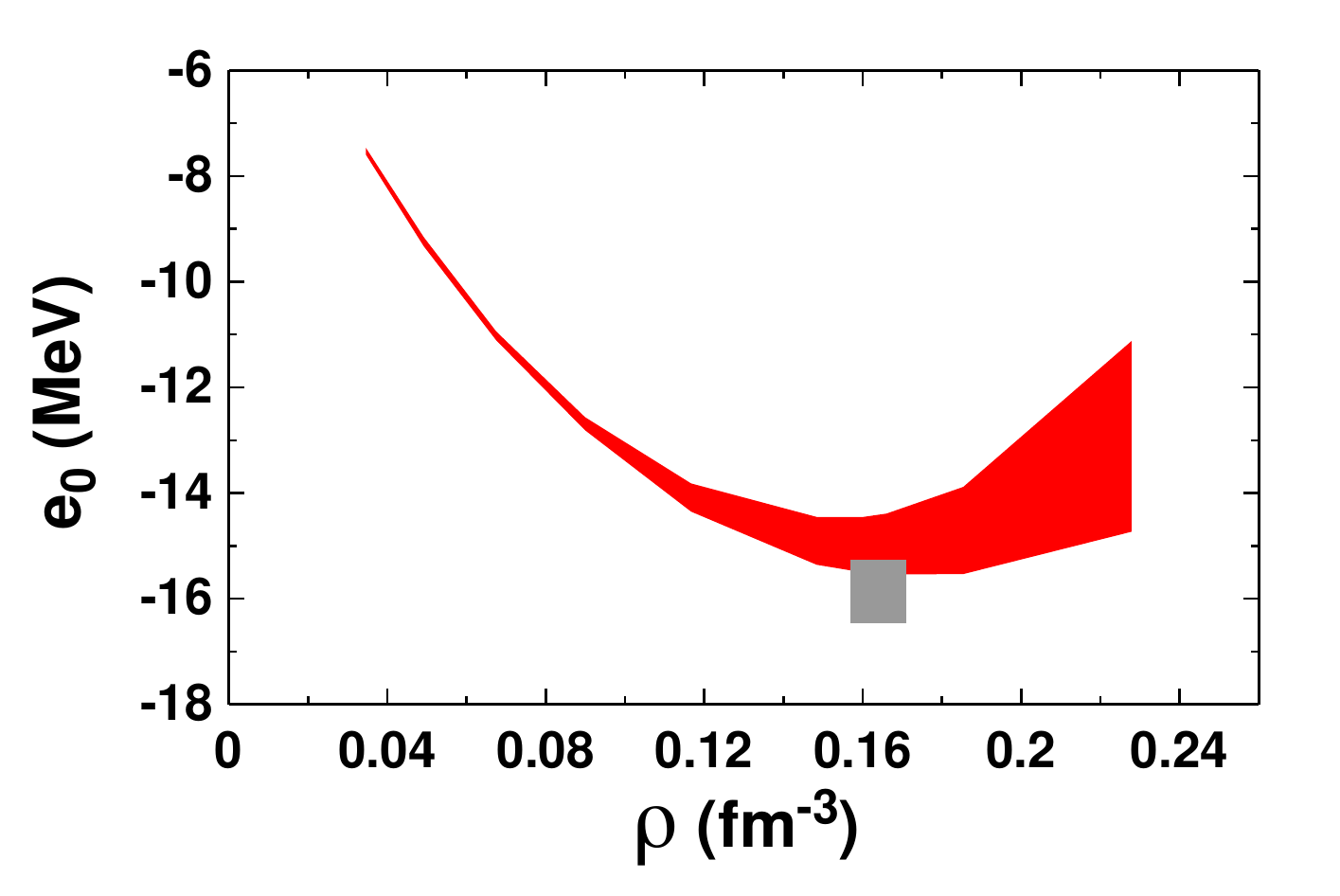}\hspace{0.01in} 
\vspace*{-0.05cm}
\caption{Energy per nucleon in SNM as a function of density at fourth order of chiral expansion. The band shows the uncertainty calculated from Eq.~\ref{err}. The grey box marks the empirical saturation point, consistent with Refs.~[22] and [48] in Ref.~\cite{DHS19}.
}
\label{band}
\end{figure*}

To evaluate the truncation error for saturation parameters using Eq.~(\ref{err}), one might define a nominal ``saturation" density, say $\rho_0$ = 0.16 fm$^{-3}$, for all orders. On the other hand,  the EoS at LO and NLO do not exhibit a saturating behavior, thus,
 it may be more meaningful to consider the actual saturation densities for the EoS which do saturate (namely, those including 3NFs), especially for the purpose of evaluating the  incompressibility, which measures the curvature of the EoS at the minimum. Estimating the truncation error at N$^3$LO as $|X_{N^3LO} - X_{N^2LO}| $, which is a pessimistic estimate, we find, for the saturation density at N$^3$LO, $\rho_0$ = (0.161 $\pm$ 0.015) fm$^{-3}$. 
Proceeding in the same way for the saturation energy and the incompressibility, we obtain, at N$^3$LO, $e(\rho_0)$ = (-14.98 $\pm$ 0.85) MeV, and $K_0$ = (216 $\pm$ 33) MeV~\cite{SNM21}.
Adopting, instead, the prescription  $|X_{N^3LO} - X_{N^2LO}|  \frac{Q}{\Lambda}$, where $Q$ is identified with the Fermi momentum at saturation density, the errors would be reduced by about 44\%.
Figure~\ref{band} displays the predictions at N$^3$LO with the uncertainty band calculated from Eq.~\ref{err}. The LECs $c_D$ and $c_E$ are equal to 0.50 and -1.25, respectively, and are taken from Ref.~\cite{DHS19}. For comparison, we display saturation properties for the final posterior predictive distribution (PPD) from Ref.~\cite{Hu+}, see Table~\ref{Hu}.

\begin{table*}[t]
\caption{Predictions for the saturation properties of SNM from the final PPD of Ref.~\cite{Hu+}. Shown are the medians, 68\% credible regions (CR) and 90\% CR.  The last column contains our predictions~\cite{SNM21}. The energy per nucleon, $E/A$, and the incompressibility, $K$, are in units of MeV. The saturation density, $\rho_0$, is in units of fm$^{-3}$.}
\label{Hu}
\begin{tabular*}{\textwidth}{@{\extracolsep{\fill}}ccccc}
\hline
\hline
 Observable & median & 68 \% CR & 90 \% CR  & our predictions \\
\hline 
\hline
$E/A$  & - 15.2 & [-16.3, -13.9] &  [17.1, -13.4]    & -14.98 $\pm$ 0.85  \\
$\rho_0$   & 0.163  & [0.147, 0.176]  & [0.140, 0.186]     & 0.161 $\pm$ 0.015                          \\ 
$K_0$       &264      & [219, 300]   &  [202, 336]  & 216 $\pm$ 33 \\
\hline
\hline
\end{tabular*}
\end{table*}

\subsection{Neutron matter and the symmetry energy}
\label{pred}

As mentioned earlier, the formation of the neutron skin in neutron-rich nuclei is a fascinating phenomenon. It is the result of excess neutrons
forced outward by the neutron-rich core of
the nucleus, which effectively amounts to a pressure gradient that moves some of the excess neutrons to the outskirts of the nucleus.
Although a small contribution to the size of the nuclear radius,
the neutron skin reveals important information about
the physics of nucleon interactions with changing density in a strongly isospin-asymmetric environment. 

The EoS of neutron-rich matter is also at the forefront of nuclear astrophysics. Neutron stars are important natural laboratories for constraining theories of the EoS for neutron-rich matter, 
to which the star mass-radius relationship is sensitive. Interest in these compact stars has increased considerably with the onset of the ``multi-messenger era" for astrophysical observations.  We will start with discussing predictions based on chiral EFT~\cite{NM21}.

\begin{figure*}[!t] 
\centering
\hspace*{-3cm}
\includegraphics[width=7.7cm]{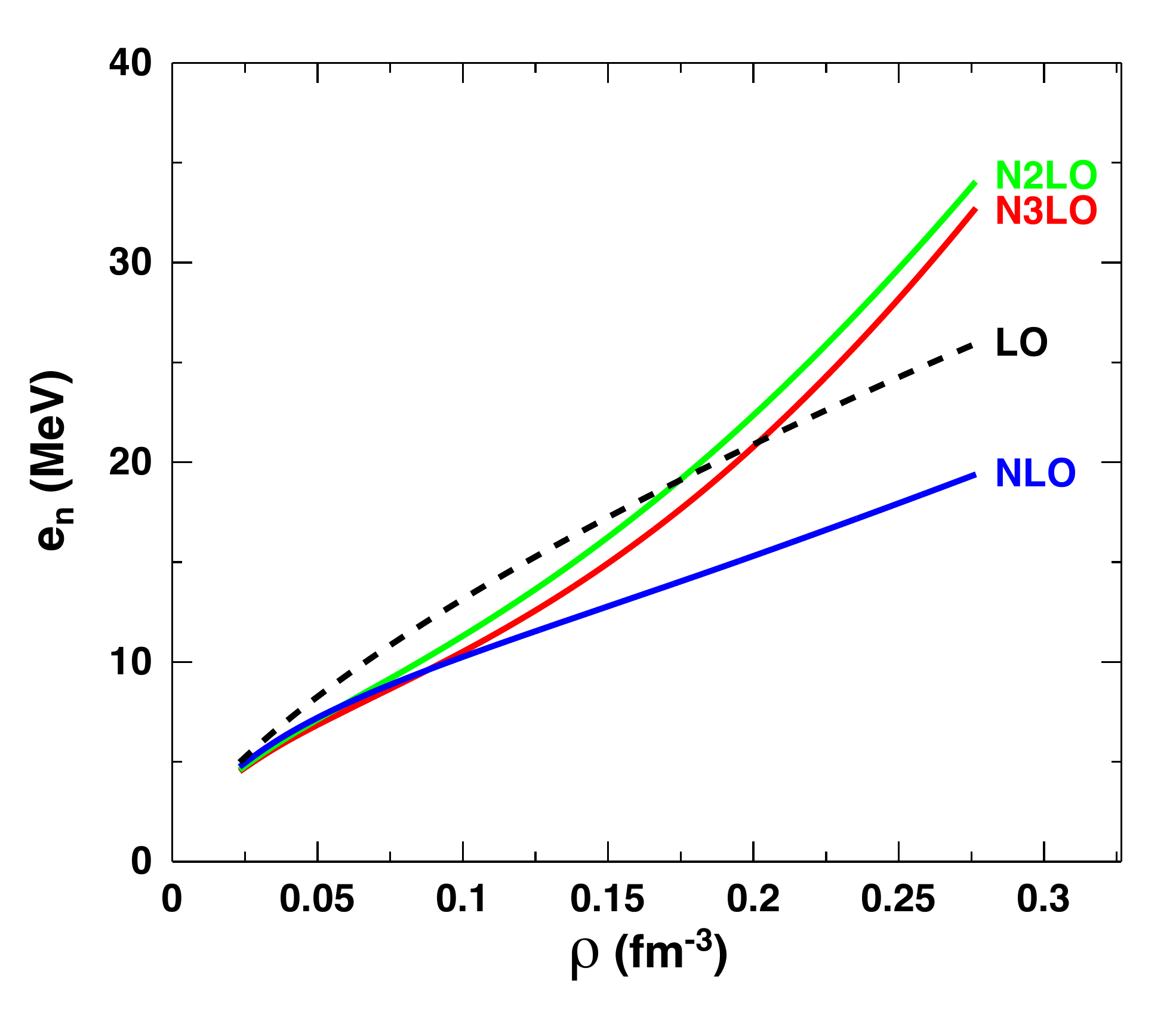}\hspace{0.01in} 
\vspace*{-0.5cm}
 \caption{(Color online) Energy per neutron in NM as a function of density, from leading order (black dash) to fourth order (solid red).
}
\label{nm_eos}
\end{figure*}   

\begin{figure*}[!t] 
\centering
\hspace*{-3cm}
\includegraphics[width=7.7cm]{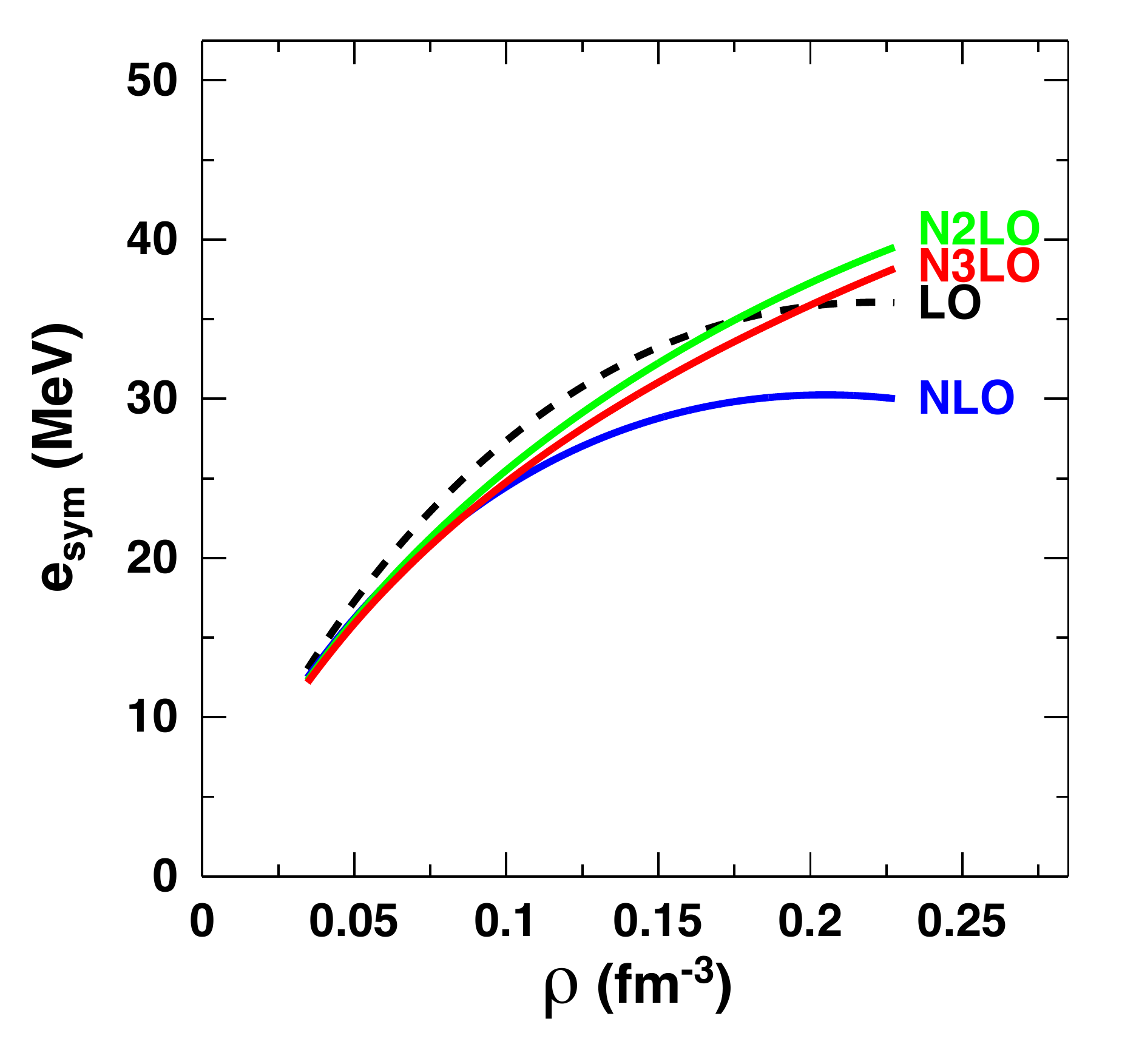}\hspace{0.01in} 
\vspace*{-0.5cm}
 \caption{(Color online)  The symmmetry energy as a function of density, from leading order (black dash) to fourth order (solid red).
}
\label{esym}
\end{figure*}

In Fig.~\ref{nm_eos}, we show our results for the energy per neutron in NM as a function of density over four chiral orders, from LO to N$^3$LO. The
large variations between NLO and N$^2$LO are mostly due to the first appearance of 3NFs at N$^2$LO. The predictions at N$^3$LO are slightly more attractive than those at N$^2$LO, in agreement with other calculations~\cite{Dri+16}.
Our NM EoS is on the soft side of the large spectrum of EoS in the literature, which is generally true for predictions based on chiral EFT, with a correspondingly soft density dependence of the symmetry energy, as shown in Fig.~\ref{esym}.

For the purpose of the present discussion, we replace the EoS of SNM with an empirical parametrization~\cite{Oya2010}, where the energy per nucleon (-16.0 MeV) and  the saturation density ($\rho_0$=0.155 fm$^{-3}$) are consistent with the traditionally cited values. We emphasize that this is just to spotlight the behavior of the energy and pressure in neutron matter and their impact on the symmetry energy {\it while the isoscalar properties remain unchanged}. Varying simultaneously isoscalar properties would obscure what we wish to highlight. Semi-empirical constraints for the symmetry energy are typically obtained by constructing families of parametrized model EoS which differ in their predictions for $L$ while remaining otherwise equivalent. Under these circumstances, one best observes the linear relation between $L$ and the neutron skin thickness. Therefore, the comparisons we explore in this section are more meaningful if we remove additional uncertainties arising, for instance, from the sensitivity of the SNM EoS to the $c_D$, $c_E$  LECs.

From a variety of phenomenological analyses, typical values for $L$ range within 70 $\pm$ 15 MeV \cite{BB_1_2016, BCZ_1_2019, ADS_1_2012, MMSY_1_2012, MSIS_1_2016, Tsang}. This is at variance with the findings from the recent PREX-II experiment~\cite{prexII}: $e_{sym}$ = (38.29 $\pm$ 4.66) MeV and $L$ = (109.56 $\pm$ 36.41) MeV, for the symmetry energy and its slope at saturation, respectively, also in disagreement with a large number of experimental measurements.

In Table~\ref{par0} we show some of our predictions at N$^3$LO with their truncation errors:
the energy per neutron and the symmetry energy at saturation, the slope parameter as defined in Eq.~(\ref{L}), and the pressure in neutron matter. As already observed, 
a softer nature is typical of chiral predictions.  A comparison with phenomenological interactions of the past, such as Argonne V18 and the UIX 3NF~\cite{APR}, can be found in Ref.~\cite{chi}.
A more recent analysis~\cite{Bays20} reports values for $e_{sym}(\rho_0)$ and $L$ of (31.7 $\pm$ 1.1) MeV and (59.8 $\pm$ 4.1) MeV, respectively.

We observe that values such as those shown in the first row of Table~\ref{par0} -- approximately $L$=(50 $\pm$ 10) MeV, and pressure at $\rho_0$ between 2 and 3 MeV/fm$^3$-- are far from those extracted from the PREX-II experiment. The PREX-II value of the pressure at $\rho_0$ is  approximately  between 3.66 MeV/fm$^3$ and 7.30 MeV/fm$^3$, an extremely stiff symmetry energy that would allow rapid cooling through direct Urca processes to proceed at unusually low values of the neutron star mass and central density~\cite{prexII}, which seems unlikely~\cite{PLPS09}. The proton fraction we obtain is close to 6\% at $\rho \approx$ 0.2 fm$^{-3}$, considerably smaller than the direct Urca threshold of approximately 1/9~\cite{urca}.

\begin{figure*}[!t] 
\centering
\hspace*{-0.5cm}
\includegraphics[width=12.5cm]{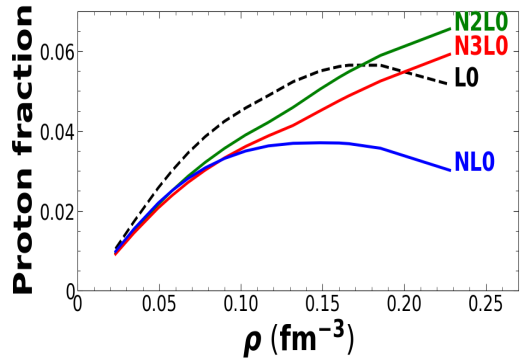}\hspace{0.01in} 
\vspace*{-9.5cm}
 \caption{(Color online)  The proton fraction in $\beta$-stable matter as a function of density, from leading order (black dash) to fourth order (solid red).
}
\label{fract}
\end{figure*}

Before we take a short detour towards neutron stars, we wish to emphasize that no theory of hadrons can describe neutron star matter from outer crust to inner core. However, the normal density region is of great relevance for the physics of neutron stars, given the sensitivity of the radius to the slope parameter $L$ for star masses around 1.4 solar masses.

We recall that the direct Urca process is the fastest cooling mechanism for neutron stars. 
It is due to thermally excited neutrons undergoing $\beta$-decay:
\begin{equation}
\label{beta-} 
n \rightarrow p + e^- + \bar{\nu}_e    \; ,
\end{equation}
while thermally excited protons undergo inverse $\beta$-decay:
\begin{equation}
\label{beta+} 
p \rightarrow n + e^+ + \nu_e    \; .
\end{equation}
The neutrinos carry away energy as they escape and the star cools very rapidly to temperatures below 10$^9$ K, at which point a minimum proton fraction (the URCA threshold mentioned above) must be present in order to preserve conservation of momentum. The proton fraction in a cold neutron star (temperature below 10$^9$ K) is determined by the symmetry energy, which is a measure of the energy change associated with variations in relative proton and neutron concentrations. 
The proton fractions we obtain in $\beta$-stable stellar matter are shown in Fig.~\ref{fract} as a function of density. The similarity of pattern with Fig.~\ref{esym} is noticeable. At typical nuclear densities, our predicted proton fraction at N$^3$LO is about 1/25.

Within the mean-field philosophy, on the other hand, one proceeds in the opposite direction. We take note, for instance, of Ref.~\cite{urca2}, where the authors construct equations of state using covariant density functional theory and explore the coupling parameter space of the isovector meson to generate a variety of models. Not surprisingly, when the constraint on $L$ from PREX-II is included, all models allow direct Urca cooling at densities as low as 1.5$\rho_0$ and within neutron stars with mass as low as one solar mass.

Back to Table~\ref{par0}, we also show predictions at some specific densities below $\rho_0$. These are the  densities identified in Ref.~\cite{Tsang} as densities that are sensitive to specific observable, as determined from the slope of the correlation in the plane of $e_{sym}(\rho_0)$  {\it vs.} $L$ obtained from the measurements of that observable. In fact, a particular slope reflects a specific density at which that observable is especially sensitive to the symmetry energy. Our {\it ab initio} predictions and the values taken from Ref.~\cite{Tsang} agree within uncertainties.
 We recall that, at $\rho$ = (2/3)$\rho_0 \approx$ 0.1 fm$^{-3}$ (an average between central and surface densities in nuclei),  the symmetry energy is well constrained by the binding energy of heavy, neutron-rich nuclei~\cite{prexII}, and therefore this density region is relevant for the purpose of correlations between $L$ and the neutron skin of $^{208}$Pb.

\begin{table*}
\caption{The energy per neutron, the symmetry energy, the slope parameter, and the pressure at N$^3$LO at various densities, $\rho$, in units of $\rho_0$=0.155 fm$^{-3}$.
$L$ is defined as in Eq.~(\ref{L}) at the specified density. The values in parentheses are taken from Ref.~\cite{Tsang}. The constraint for $L$ in the third row ($\rho =0.67 \rho_0$) is given for  $\rho$=0.1 fm$^{-3}$. The constraint at $\rho =0.31 \rho_0$ is from Ref.~\cite{ZZ15}. }
\label{par0}
\centering
\begin{tabular*}{\textwidth}{@{\extracolsep{\fill}}ccccc}
\hline
\hline
 $\rho$ $ (\rho_0) $ & $ \frac{E}{N}(\rho)$ (MeV) & $e_{sym}(\rho)$ (MeV)  & $L (\rho)$(MeV)  &  $P_{NM}(\rho)$ (MeV/fm$^3$)  \\
\hline
\hline
 1 & 15.56  $\pm$ 1.10& 31.57 $\pm$ 1.53 (33.3 $\pm$ 1.3)  & 49.58 $\pm$ 8.47 (59.6 $\pm$ 22.1) & 2.562  $\pm$ 0.438 (3.2 $\pm$ 1.2)  \\  
0.72 (0.72 $\pm$ 0.01) & 11.52  $\pm$ 0.43 & 26.46 $\pm$ 0.82 (25.4 $\pm$ 1.1) & 44.91 $\pm$ 3.40 & 1.05  $\pm$ 0.13  \\  
 0.67 (0.66 $\pm$ 0.04) & 10.81  $\pm$ 0.41 & 25.25 $\pm$ 0.72 (25.5 $\pm$ 1.1) & 44.65 $\pm$ 3.23 (53.1 $\pm$ 6.1) & 0.859  $\pm$ 0.120  \\  
0.63 (0.63 $\pm$ 0.03) & 10.39  $\pm$ 0.41 & 24.47  $\pm$ 0.66 (24.7 $\pm$ 0.8) & 43.81 $\pm$ 3.11 & 0.748  $\pm$ 0.116  \\  
0.31 (0.31 $\pm$ 0.03) &   6.715 $\pm$ 0.086  &15.43 $\pm$ 0.12  (15.9 $\pm$ 1.0) &  32.35 $\pm$ 0.55    &     0.174 $\pm$ 0.008                            \\ 
0.21 (0.22 $\pm$ 0.07) & 5.472  $\pm$  0.039 & 11.73 $\pm$ 0.05 (10.1 $\pm$ 1.0) & 27.57 $\pm$ 0.11 & 0.106  $\pm$ 0.002  \\  
\hline
\hline
\end{tabular*}
\end{table*} 

In summary, a range for $L$ between 45 and 65 MeV can be taken as typical for state-of-the-art nuclear theory predictions.

We close this section with a few comments on the symmetry energy at higher densities. Heavy-ion collisions are typical experiments used to probe higher densities. For instance, directed and elliptic flows of neutrons and light-charged particles were measured for the reaction $^{197}$Au + $^{197}$Au at an incident energy of 400 MeV/nucleon incident energy in the ASY-EOS experiment at the GSI laboratory~\cite{R_etal_1_2016}. The densities probed are shown to reach beyond twice saturation. For both low and high density, one must be cautious with the interpretation of constraints extracted from the measured observables, which may be model dependent.

Moving to even higher densities, the composition of a neutron star core will remain largely unknown unless reliable constraints on the symmetry energy at high densities become available. The minimum mass for direct Urca cooling may be a suitable constraint for the density dependence of the symmetry energy~\cite{Malik22} above approximately 3$\rho_0$, which is found to be
strongly correlated with the neutron star mass at which the onset of direct Urca neutrino cooling takes place in the core. The analysis from Ref.~\cite{Malik22_2} finds that the values of tidal
deformability and radius for a 1.4 $M_{\odot}$ neutron star
are correlated with the pressure in $\beta$-equilibrated
matter at about 2$\rho_0$.

From the theoretical standpoint, 
densities around twice saturation density or higher are outside the reach of current {\it ab initio} calculations. Nevertheless, the normal density regime is of far reaching importance for higher density features.

\subsection{Impact on the neutron skin}
\label{skin} 

Relating the nucleus spatial extension as directly as possible to the microscopic EoS is best achieved by means of the droplet model. This also allows applications to heavy nuclei, which may be outside the reach of {\it ab initio} methods.

Recently, we calculated the neutron skin~\cite{FS_2022},
\begin{equation}
S = <r^2>_n^{1/2} -  <r^2>_p^{1/2} \; ,
\end{equation}
of $^{208}$Pb using droplet model expressions where the symmetry energy and its density slope at saturation appear explicitely~\cite{FS_2022},
 see Table~\ref{tab1}. We emphasize that the simple droplet model is being used because we can directly input the values of $J$ and $L$ -- the table entries relative to one another should be the focal point.

\begin{table*}
\caption{ The neutron skin of $^{208}$Pb, $S$, calculated as described in Ref.~\cite{FS_2022} using the specified symmetry energy, $J$, and its slope at saturation, $L$. 
}
\label{tab1}
\begin{tabular*}{\textwidth}{@{\extracolsep{\fill}}cccc}
\hline
\hline
 $J$ (MeV)      & $L$ (MeV)  & $S$ (fm)  & source for $J$, $L$   \\
\hline    
\hline 
 31.3 $\pm$ 0.8  &  52.6 $\pm$ 4.0  & [0.13, 0.17] & \cite{Universe}   \\
 (31.1, 32.5)   &  [44.8, 56.2]  & [0.12, 0.17] &  \cite{DHS19}   \\
(28, 35)  &  [20, 72]  &     [0.078, 0.20] &     \cite{DHW21}    \\
(27, 43)                & [7.17, 135]  &  [0.055, 0.28]           &  \cite{LH19}    \\     
 38.29  $\pm$ 4.66  &  109.56  $\pm$ 36.41  &  [0.17, 0.31] &  \cite{prexII}   \\              
\hline
\hline
\end{tabular*}
\end{table*}

The first three entries in Table~\ref{tab1} are obtained from EoS based on chiral EFT, with chiral two- and three-nucleon interactions at N$^3$LO. One can see that they are relatively soft, cover a narrow range, and are in good agreement with one another. Consistent with earlier discussions,  the corresponding neutron skins are relatively small. Most recently,
ab initio predictions for the neutron skin of $^{208}$Pb have become
available~\cite{Hu+}. The reported range is between 0.14
and 0.20 fm, smaller than the values extracted from parity violating
electron scattering. 
The fourth line correspond to an analysis based on current constraints from nuclear theory and experiment. 
In Ref.~\cite{LH19}, the authors utilized 48 phenomenological models, both relativistic mean field and Skyrme Hartree-Fock. 
The last line shows the values of $J$ and $L$ from the recent PREX II experiment. Note that the reported value for the skin of $^{208}$Pb  in Ref.~\cite{prexII} is (0.283 $\pm$ 0.071) fm, giving a range between 0.21 and 0.35 fm.
 
 Clearly, a much larger range
for the neutron skins can be obtained with phenomenological
interactions, both relativistic and nonrelativistic mean-field
models, including values that are consistent with the findings of PREX
II. This is not surprising, because much larger variations
in $L$ can be generated through parametrizations of mean-field models.  
On the other hand, the realistic nature of few-nucleon forces should be 
 preserved in heavier systems, both on principle grounds -- based on the ab initio philosophy -- and
in practice, see below.  
For these reasons,
mean-field models, while remaining an important tool to explore
sensitivities and correlations, lack the predictive power
needed to shed light on open questions in {\it ab initio} nuclear structure.

\begin{figure*}[!t] 
\centering
\hspace*{-3cm}
\includegraphics[width=11.0cm]{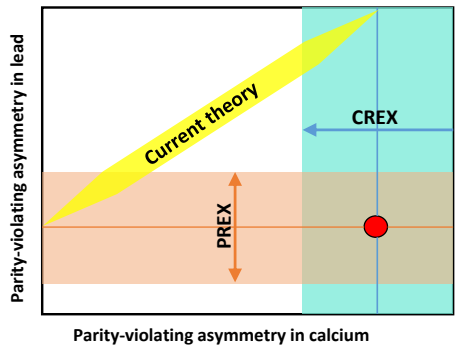}\hspace{0.01in} 
\vspace*{-8.5cm}
 \caption{(Color online) Discrepancy between the CREX and PREX measurements and theoretical predictions. See text and Ref.~\cite{msu} for more details.
}
\label{PCrex}
\end{figure*}    

We conclude this section pointing to the interesting discussion in Ref.~\cite{crex}, where the authors 
extended their previous analysis of the PREX experiment to recent measurements of parity-violating asymmetry in $^{48}$Ca (CREX experiment). 
 The study included the static electric dipole polarizability, an observable that is expected to correlate with the neutron skin. We fully agree that a critical search is needed for the limitations of interactions currently used in {\it ab initio} calculations and other sources of experimental error.
Figure~\ref{PCrex}, generated from information extracted from the figure in Ref.~\cite{msu}, 
shows the discrepancy between CREX and PREX measurements and current theoretical predictions. The lines mark the mean values of the measured asymmetry in lead and calcium, respectively, and the bands represent the errors. Theoretical predictions should be, but are not, at the intersection of CREX and PREX results, marked by the red dot.

\section{Impact of the isovector part of the free-space $NN$ force}
\label{1s0}

In the impressive analysis in Ref.~\cite{Hu+}, 10$^9$ nuclear force parametrizations consistent with chiral EFT are examined. Employing state-of-the-art statistical methods and computational technology, the authors are able to make quantitative predictions for bulk properties and skin thickness of $^{208}$Pb. In the process, they find a strong correlation between $L$ and the $^1S_0$ phase shift at laboratory energies around 50 MeV.

Here, we perform a single-shot calculation which serves as a simple and transparent test of the impact of the isovector part of the free-space $NN$ force in neutron matter. We consider 2NFs only.

\begin{figure*}[!t] 
\centering
\hspace*{-3cm}
\includegraphics[width=9.0cm]{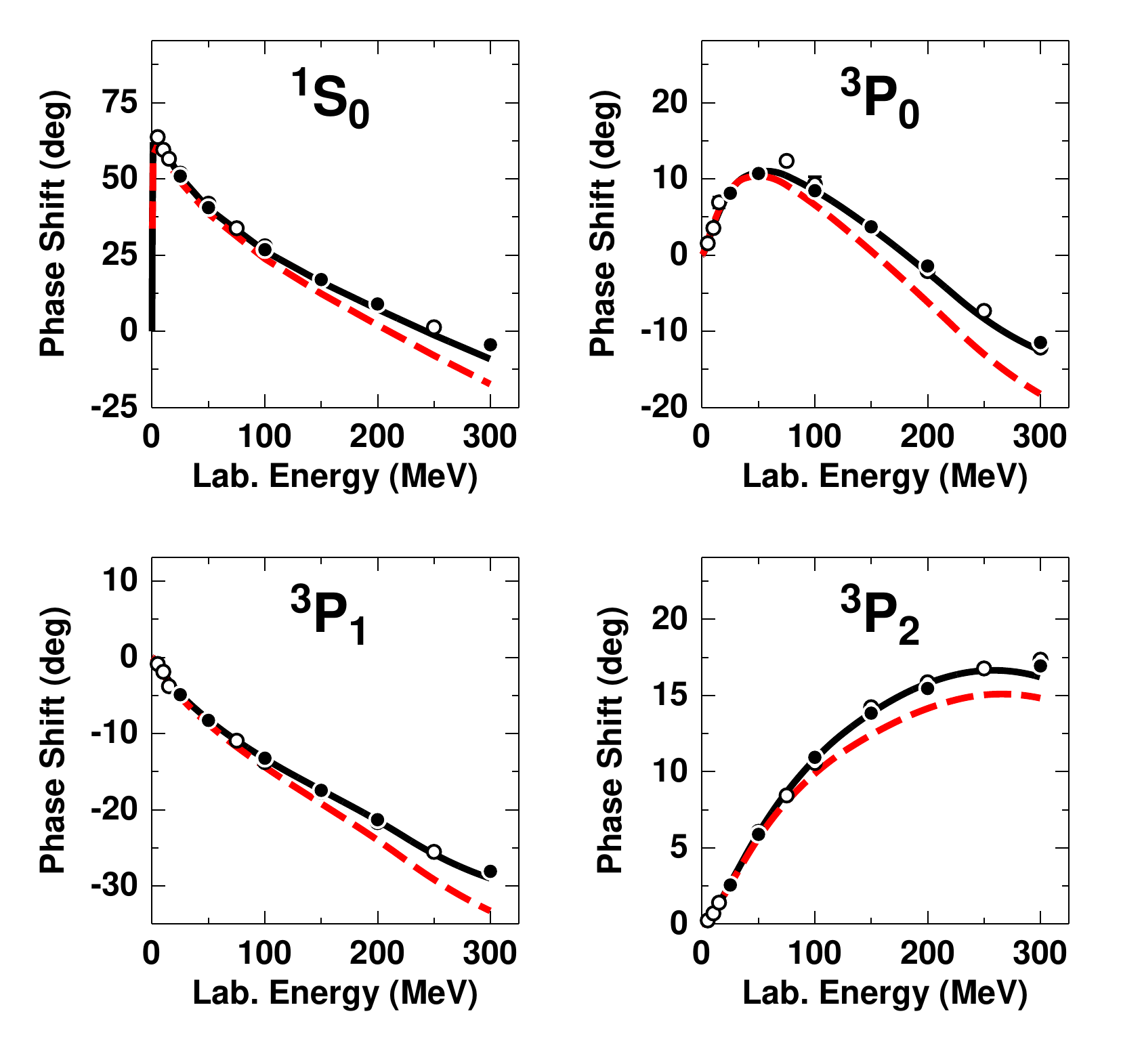}\hspace{0.01in} 
\vspace*{-0.2cm}
 \caption{(Color online) Phase shifts for selected isospin-1 partial waves as a function of the laboratory energy. Solid black: original potential; Red dash: modified potential.
}
\label{bad3}
\end{figure*}

For the purpose of this test, we made a version of the N$^3$LO(450) potential where the fit of isospin-1 partial waves is deteriorated as compared to the original potential, see Fig.~\ref{bad3}. This is accomplished by adjusting two LECs in $^1S_0$ channel  and one LEC for each of the $P$-waves (all changes are on average between 2 and 10\%), while keeping the scattering length at its correct value. Although not dramatic, the impact on the phase shifts is considerable, especially above 100 MeV.

\begin{figure*}[!t] 
\centering
\hspace*{-1.0cm}
\includegraphics[width=11.5cm]{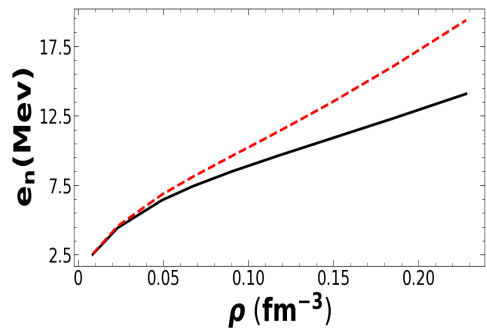}\hspace{0.01in} 
\vspace*{-8.5cm}
 \caption{(Color online) Energy per neutron as a function of NM density. Solid black: original potential; Red dash: modified potential.
}
\label{enm_bad}
\end{figure*}

We then calculate the EoS of NM with the modified potential, shown in Fig.~\ref{enm_bad} along with the original N$^3$LO predictions (2NF only).
As we don't include 3NFs, these values are not realistic, but we are interested in highlighting differences due to the 2NF. Around saturation, the energy moves up by about 25\%, while the (very sensitive) slope and closely related pressure increase by a factor of 1.75. This comparison is shown in Table~\ref{tab_bad3}.
In other words, the ``modified" phase shifts are not disastrous, but the slope of the NM EoS changes dramatically.  
 Notice that our modified interaction is only very little different from the original interaction below 100, confirming extreme sensitivity of the neutron matter slope (and thus the slope of the symmetry energy) to the description of the isovector component of the $NN$ interaction. Of course, a more ``quantitative" test would require the inclusion of 3NFs, readjusted for consistency with the 2NF modifications. On the other hand, this simple exercise suffices to demonstrate that relaxing or abandoning the constraint of free-space $NN$ data can produce dramatic changes in $L$ (and thus the neutron skin). 

The freedom to modify a model in such a way that isovector properties (such as the slope of the NM EoS) vary while retaining good fits to nuclei and nuclear matter, is the mechanism that generates the popular correlations from mean-field models. However, free-space $NN$ data do not enter in this picture, contrary to the basic principle of {\it ab initio} predictions. Also, variations within the model parameter space are generally applied to predefined analytical expressions, such as power laws of the density.

\begin{table*}
\caption{ The energy per neutron, its slope, and the pressure at a density of 0.155 fm$^{-3}$ with the originl and the modified potentials. Only the 2NF is included.
}
\label{tab_bad3}
\begin{tabular*}{\textwidth}{@{\extracolsep{\fill}}ccc}
\hline
\hline
  calculated quantity      & N$^3$LO(450)  & modified potential   \\
\hline    
\hline 
 $e_n(\rho_0)$ (MeV)  &  11.11  & 13.88   \\
 $ \Big (\frac{\partial e_n(\rho)}{\partial \rho} \Big )_{\rho_0} $ (MeV/fm$^{-3}$) & 39.79 & 70.03   \\
 $ P(\rho_0)$ (MeV/fm$^{3}$)  & 0.956 & 1.68    \\  
\hline
\hline
\end{tabular*}
\end{table*}

\section{Conclusions and future perspective: Where do we go from here? }
\label{conc}

We performed an analysis of existing literature addressing the nuclear symmetry energy, with the objective to identify and discuss current gaps or problems, and provide recommendations for future research.

Nuclear theory has come a long
way from the days of one-boson-exchange nucleon-nucleon
potentials and attempts to incorporate selected 3NFs with no clear scheme or guidance. As for phenomenological interactions
(such as relativistic and non-relativistic mean-field models), they
are a very useful tool to probe sensitivities and explore correlations
but, {\it by their very nature}, cannot address important
questions in {\it ab initio} nuclear structure.

 Thanks to continuous
progress in nuclear theory, one is now able to construct nuclear
forces in a systematic and internally consistent manner.
The order-by-order structure inherent to chiral EFT allows
to explore the importance of different contributions from
few-nucleon forces as they naturally emerge at each order. For a better understanding of intriguing systems such as neutron skins
and neutron stars, it is important to build on that progress.
Predictions from state-of-the-art nuclear theory favor a softer density dependence of the symmetry energy -- on the low-to-medium end of what is considered an acceptable range -- and, consistently, smaller values of the neutron skin and the radius of the average-mass neutron star.

On the experimental side, the symmetry energy parameters
that drive the neutron skin (and the radius of low to medium mass neutron stars) are not measured directly, but rather
extracted from measurements of suitable observables. While
electroweak (EW) methods avoid the uncertainty inherent to
the use of hadronic probes, the weakness of the signal seems
to generate large errors -- the measured observable is the very small left-right asymmetry in weak electron scattering off nuclei. This may interfere with the ability of
the result to provide a benchmark.

 It can be demonstrated, and should be espected on fundamental grounds, that accurate reproduction of low-energy $NN$ data is an important requirement for realistic predictions of many-body systems. Ignoring or weakening that constraint takes us back to mean-field approaches and thus is not progress. 

Whether one chooses to call it ``{\it tension}"~\cite{Hu+} or  ``{\it irreconcilable differences}"~\cite{FS_2022}, large pressure in neutron matter at saturation and large neutron skins are inconsistent with 
essentially all state-of-the-art predictions.
We suggest that the way forward is for the low-energy experimental and theory communities to work closely. Theorists should continue to confront systematic problems in nuclear structure, critically examining the interactions currently used in {\it ab initio}:/grap calculations. Claims that most recent results from EW scattering represent a benchmark are premature.

\section*{Acknowledgments}
This work was supported by 
the U.S. Department of Energy, Office of Science, Office of Basic Energy Sciences, under Award Number DE-FG02-03ER41270.


\begin{references}   
  
\bibitem{frib} https://frib.msu.edu

\bibitem{skyrm1} Zahed, I; Broen, G E. The Skyrme model. {\it Phys. Rept.} {\bf 1986}, 142, 1.

\bibitem{skyrm2} Guo-qiang, L. Skyrme forces and their Applications in Low Energy Nuclear Physics. {\it Commun. Theor. Phys.} {\bf 1990}, 13, 457.

\bibitem{skyrm3} Bender, M; Heenen, P-H; Reinhard, P-G. Self-consistent mean-field models for nuclear structure. {\it Rev. Mod. Phys.} {\bf 2003}, 75, 121.

\bibitem{rmf1} Waletcka, J D. A Theory of highly condensed matter. {\it Annals Phys.} {\bf 1974}, 83, 491.

\bibitem{rmf2} Serot, B D; Waletcka, J D. The Relativistic Nuclear Many Body Problem. {\it Adv. Nucl. Phys.} {\bf 1986}, 16, 1.

\bibitem{rmf3} Serot, B D; Waletcka, J D. Recent Progress in Quantum Hadrodynamics. {\it Int. J. Mod. Phys.} {\bf 1997}, E6, 515.

\bibitem{B_1_2000}
Brown, B A. Neutron Radii in Nuclei and the Neutron Equation of State.  {\it Phys. Rev. Lett.} {\bf 2000}, 85, 5296.

\bibitem{PF_1_2019}
 Piekarewicz, J; Fattoyev, F. Neutron-rich matter in heaven and on Earth. {\it Physics Today} {\bf 2019}, 72, 7, 30.


\bibitem{SDLD_1_2015} Santos, B M; Dutra, M. Lourenco, O; Delfino, A.
 Correlations between the nuclear matter symmetry energy, its slope, and curvature.
 {\it Jour. Phys. Conf. Ser.} {\bf 2015}, 630, 012033.


\bibitem{FHPS_1_2010} Fattoyev, F J; Horowitz, C J; Piekarewicz, J; Shen, G. Relativistic effective interaction for nuclei, giant resonances, and neutron stars {\it Phys. Rev. C} {\bf 2010}, 82, 055803.


\bibitem{MCVW_1_2011} Roca-Maza, X; Centelles, M; Vinas, X; Warda, M. Neutron Skin of $^{208}$Pb,
Nuclear Symmetry Energy, and the Parity Radius Experiment.{\it  Phys. Rev. Lett.} {\bf 2011}, 106, 252501.


\bibitem{MADSCV_1_2017} Mondal, C; Agrawal, B K; De, J N; Samaddar, S K; Centelles, M; Vinas, X.
Searching for a universal correlation among symmetry energy parameters.
Proc. DAE Symp. {\it Nucl. Phys.} {\bf 2017} 62, 72.

\bibitem{TLOK17} Tews, I; Lattimer, J M; Ohnishi, A; Kolomeitsev, E E. Symmetry Parameter Constraints from a Lower Bound on Neutron-matter Energy.  {\it  Astrophys. J.} {\bf 2017}, 848, 105.

\bibitem{MADS_1_2018} Mondal, C; Agrawal, B K; De, J N; Samaddar, S K.
Correlations among symmetry energy elements in Skyrme models.
 {\it Int. Jour. Mod. Phys. E} {\bf 2018}, 27, 1850078.


\bibitem{TRRSWM_1_2018} Tong, H; Ren, X-L; Ring, P; Shen, S-H; Wang, S-B; Meng, J. Relativistic Brueckner-Hartree-Fock theory in nuclear matter without the average momentum approximation.
{\it Phys. Rev. C} {\bf 2018}, 98, 054302. 

\bibitem{HY_1_2018} Holt, J W; Lim,Y.
Universal correlations in the nuclear symmetry energy, slope parameter, and curvature.
 {\it Phys. Lett. B} {\bf 2018}, 784, 77.  

\bibitem{ADSCS_1_2013} Agrawal, B K; De, J N; Samaddar, S K; Colo, G; Sulaksono, A.
Constraining the density dependence of symmetry energy from nuclear masses.
 {\it Phys. Rev. C} {\bf 2013}, 87, 051306. 

\bibitem{VCRW_1_2014} Vinas, X; Centelles, M; Roca-Maza, X; Warda, M.
Density dependence of the nuclear symmetry energy from measurements of neutron radii in nuclei.
 {\it AIP Conf. Proc.} {\bf 2014}, 1606, 256.


\bibitem{ADS_1_2012}  Agrawal, B K; De, J N; Samaddar, S K.
Determining the density content of symmetry energy and neutron skin: an empirical approach.
{\it Phys. Rev. Lett.} {\bf 2012}, 109, 262501. 

\bibitem{Tsang+09} Tsang, M B; Y. Zhang, Y; Danielewicz, P; Famiano, M; Z. Li,  Z; Lynch, W G {\it  et al.}
Constraints on the density dependence of the symmetry energy.
 {\it Phys. Rev. Lett.} {\bf 2009}, 102, 122701. 

\bibitem{Tsang+12} Tsang, M B {\it et al.}  Constraints on the symmetry energy and neutron skins from experiments and theory.
{\it Phys. Rev. C} {\bf 2012}, 86, 015803.  

\bibitem{LL13} Lattimer, J M; Lim, Y. Constraining the symmetry parameters of the nuclear interaction.  {\it Astrophys. J.} {\bf 2013}, 771, 51.  

\bibitem{Kort+10} Kortelainen, M; Lesinski, T; Moré, J; Nazarewicz, W; Sarich, J; Schunck, N; Stoitsov, M V; Wild, S.  Nuclear energy density optimization.
 {\it Phys. Rev. C} {\bf 2010}, 82, 024313. 

\bibitem{DL_1_2014} Danielewicz, P; Lee, J.
Symmetry energy II: isobaric analog states.
 {\it Nucl. Phys. A} {\bf 2014}, 922, 1. 


\bibitem{Roca+15} Roca-Maza, X; Viñas, X; Centelles, M; Agrawal, B K; Colò, G; Paar, N; Piekarewicz, J; Vretenar, D. Neutron skin thickness from the measured electric dipole polarizability in $^{68}$Ni, $^{120}$Sn, and $^{208}$Pb.
 {\it Phys. Rev. C} {\bf 2015}, 92, 064304. 


\bibitem{Tam+11} Tamii A {\it et al.}  Complete Electric Dipole Response and the Neutron Skin in $^{208}$Pb.
{\it Phys. Rev. Lett. } {\bf 2011}, 107, 062502.

\bibitem{Brown13} Brown, B A. Constraints on the Skyrme Equations of State from Properties of Doubly Magic Nuclei. {\it Phys. Rev. Lett.} {bf 2013}, 111, 232502. 


\bibitem{R_etal_1_2016} Russotto, P; Gannon, S; Kupny, S; Lasko, P; Acosta, L; Adamczyk, M. {\it et al.}
Results of the ASY-EOS experiment at GSI: the symmetry energy at suprasaturation density.
{\it Phys. Rev. C}  {\bf 2016}, 94, 034608. 


\bibitem{R_etal_1_2011} Russotto, P;  Wu, P Z; Zoric, M; Chartier, M; Leifels, Y; R.C. Lemmon, R C {\it et al.}
Symmetry energy from elliptic flow in $^{197}$Au + $^{197}$Au.
 {\it Phys. Lett. B} {\bf 2011}, 697, 471.  

\bibitem{Wei90} Weinberg, S. Nuclear forces from chiral lagrangians.
{\it Phys. Lett. B} {\bf 1990}, 251, 288. 

\bibitem{Wei92} 
Weinberg, S. Three-body interactions among nucleons and pions. {\it Phys. Lett. B} {bf 1992}, 295, 114.

\bibitem{prexII} Reed, B T; Fattoyev,  F J; Horowitz, C J; Piekarewicz, J. Implications of PREX-II on the equation of state of neutron-rich matter.
 {\it Phys. Rev. Lett.} {\bf 2021}, 126, 172503. 

\bibitem{Universe} Sammarruca, F;  Millerson, R. The Equation of State of Neutron-Rich Matter at Fourth Order
of Chiral Effective Field Theory and the Radius of a
Medium-Mass Neutron Star.  {\it Universe} {\bf 2022}, 8, 133.

\bibitem{LR1} Alam, N; Agrawal, B K; Fortin, M; Pais, H; Providencia, C; Raduta, Ad R; Sulaksono, A. Strong correlations of neutron star radii with the slopes of nuclear matter incompressibility and symmetry energy at saturation. {\it Phys. Rev. C} {\bf 2016}, 94, 052801.

\bibitem{Cor+} Coraggio, L; Holt, J W; Itaco, N; Machleidt, R; Marcucci, L E; Sammarruca, F. Nuclear-matter equation of state with consistent two- and three-body perturbative chiral interactions.
 {\it Phys. Rev. C} {\bf 2014}, 89, 044321.

\bibitem{HF} Carbone, A; Cipollone A; Barbieri, C; Rios, A; Polls, A.       {\it Phys. Rev. C} {\bf 2013}, 88, 054326. 


\bibitem{ME_1_2011} Machleidt, R; Entem, D R.  {\it Physics Reports} {\bf 2011}, 503, 1.   

\bibitem{Epel15} Epelbaum, E; Krebs, H; Mei\ss ner,  U-G.        {\it Eur. Phys. J. A} {\bf 2015}, 51, 53. 

\bibitem{EMN17} Entem, D R; Machleidt, R; Nosyk, Y.
{\it Phys. Rev. C} {\bf 2017}, 96, 024004.

\bibitem{Hofe+} Hoferichter, M; Ruiz de Elvira, J; Kubis, B; Meissner, U-G.  {\it Phys. Rev. Lett. } {\bf 2015}, 115, 192301. {\bf 115}, 19230; Phys. Rep. {\bf 625}, 1 (2016). 
\bibitem{Hofe++} Hoferichter, M; Ruiz de Elvira, J; Kubis, B; Meissner, U-G.  {\it Phys. Rep.} {\bf 2016}, 625, 1.


\bibitem{Hop17}
  Hoppe, J; Drischler, C; Furnstahl, R J; Hebeler, K; Schwenk, A.
 Weinberg eigenvalues for chiral nucleon-nucleon interactions.
  {\it Phys. Rev. C} {\bf 2017}, 96, 054002.   

\bibitem{DHS19} Drischler, C; Hebeler, K; Schwenk, A.
  Chiral interactions up to next-to-next-to-next-to-leading order and nuclear saturation.
  {\it Phys. Rev. Lett.} {\bf 2019}, 122, 042501.   

 \bibitem{Epe02} 
Epelbaum, E; Nogga, A; Gl\"ockle, W; Kamada, H;  U.-G. Mei\ss ner, U-G; Witala, H.
Three nucleon forces from chiral effective field theory.
{\it Phys. Rev. C} {\bf 2002}, 66, 064001.   

\bibitem{holt09} Holt, J W; Kaiser, N; Weise, W. Chiral three-nucleon interaction and the $^{14}$C-dating $\beta$ decay.
 {\it Phys. Rev. C} {\bf 2009}, 79, 054331.


\bibitem{holt10}  Holt, J W; Kaiser, N; Weise, W. Density-dependent effective nucleon-nucleon interaction from chiral three-nucleon forces.  {\it Phys. Rev. C} {\bf 2010}, 81, 024002.


\bibitem{HS10} Hebeler, K; Schwenk, A. Chiral three-nucleon forces and neutron matter. {\it Phys. Rev. C} {\bf 2010}, 82, 014314.

\bibitem{Ber08} Bernard, V; Epelbaum, E; Krebs, H; Mei\ss  ner, U-G. Subleading contributions to the chiral three-nucleon force: Long-range terms.  {\it  Phys. Rev. C} {\bf 2008}, 77,  064004.

\bibitem{Ber11} Bernard, V; Epelbaum, E; Krebs, H; Mei\ss  ner, U-G. Subleading contributions to the chiral three-nucleon force. II. Short-range terms and relativistic corrections.  {\it  Phys. Rev. C} {\bf 2011}, 84,  054001.


\bibitem{Kais19}  Kaiser, N; Singh, B. Density-dependent NN interaction from subleading chiral three-nucleon forces: Long-range terms.  {\it Phys. Rev. C} {\bf 2019}, 100, 014002.   
\bibitem{Kais20} Kaiser, N. Density-dependent nn-potential from subleading chiral three-neutron forces: Long-range terms. arXiv:2010.02739v4 [nucl-th].

\bibitem{Kais18} Kaiser, N; Niessner, V. Density-dependent 
NN interaction from subleading chiral 3N forces: Short-range terms and relativistic corrections. {\it Phys. Rev. C} {\bf 2018}, 98, 054002.

\bibitem{Treur} Treuer, L. Density-Dependent Neutron-Neutron Interaction from Subleading Chiral Three-Neutron Forces.
arXiv:  2009.11104 [nucl-th].

\bibitem{Suppl} Supplemental Material for Ref.~\cite{DHS19}. http://link.aps.org/supplementsl/10.1103/PhysRevLett.122.042501.

\bibitem{fake} Nosyk, Y; Entem, D R; Machleidt, R. Nucleon-nucleon potentials from $\Delta$-full chiral effective-field-theory and implications. {\it Phys. Rev. C} {\bf 2021}, 104, 054001.

\bibitem{SNM21} Sammarruca, F; Millerson, R. Overview of symmetric nuclear matter properties from chiral interactions up to fourth order of the chiral expansion. {\it Phys. Rev. C} {\bf 2021}, 104, 064312. 

\bibitem{Hu+} Hu, B; Jiang, W; Miyagi, T;  Sun, Z; Ekstr{\" o}m, A; Forssen, C;
Hagen, G; Holt, J D; Papenbrock, T;  Stroberg, S R; 
Vernon, I. Ab initio predictions link the neutron skin of $^{208}$Pb to nuclear forces. {\it Nature Physics} {\bf 2022}, 18, 1196-1200.

\bibitem{NM21} Sammarruca, F; Millerson, R. Analysis of the neutron matter equation of state and the symmetry energy up to fourth order of chiral effective field theory.
{\it Phys. Rev. C} {\bf 2021}, 104, 034308. 


\bibitem{Dri+16} Drischler, C; Carbone, A; Hebeler, K; Schwenk, A. Neutron matter from chiral two- and three-nucleon calculations up to N3LO.
 {\it Phys. Rev. C } {\bf 2016}, 94, 054307.

\bibitem{Oya2010} Oyamatsu, K; lida, K; Koura, H. Neutron drip line and the equation of state of nuclear matter. {\it Phys. Rev. C} {\bf 2010}, 82, 027301.


 \bibitem{BB_1_2016}  Baldo, M; Burgio, G F. The Nuclear Symmetry Energy.  {\it Prog. Part. Nucl. Phys.} {\bf 2016}, 91, 203. 


\bibitem{BCZ_1_2019} Burrello, S; Colonna, M; Zheng, H.
 The Symmetry Energy of the Nuclear EoS: A Study of Collective Motion and Low-Energy Reaction Dynamics in Semiclassical Approaches.
{\it Front. Phys.} {\bf 2019}, 7, 53.



\bibitem{MMSY_1_2012} Moeller, P; Myers, W D; Sagawa, H; Yoshida, S.
New Finite-Range Droplet Mass Model and Equation-of-State Parameters.
 {\it Phys. Rev. Lett.} {\bf 2012}, 108, 052501. 

\bibitem{MSIS_1_2016} Moeller, P; Sierk, A J; Ichikawa, T; Sagawa, H.
Nuclear ground-state masses and deformations (FRDM) 2012.
 {\it Atomic Data and Nuclear Tables} {\bf 2016}, 109, 1. 

\bibitem{Tsang} Lynch, W G; Tsang, M B. Decoding the density dependence of the nuclear symmetry energy. {\it Phys. Lett. B} {\bf 2022}, 830, 137098. 

\bibitem{ZZ15} Zhang, Z; Chen, L-W.  Electric dipole polarizability in $^{208}$Pb
 as a probe of the symmetry energy and neutron matter around $\rho_0/3$.
{\it Phys. Rev. C} {\bf 2015}, 92, 031301(R).

\bibitem{APR} Akmal, A; Pandharipande, V R; Ravenhall, D G. Equation of state of nucleon matter and neutron star structure. {\it Phys. Rev. C} {\bf 1998}, 58, 1804.

\bibitem{chi} Drischler, C; Soma, V; Schwenk, A. Microscopic calculations and energy expansions for neutron-rich matter.  {\it Phys. Rev. C} {\bf 2014}, 89, 025806.

\bibitem{Bays20} Drischler, C; Furnstahl, R J; Melendez, J A; Phillips, D R. How Well Do We Know the Neutron-Matter Equation of State at the Densities Inside Neutron Stars? A Bayesian Approach with Correlated Uncertainties.  {\it Phys. Rev. Lett.} {\bf 2020}, 125, 202702. 



\bibitem{PLPS09} Page, D; Lattimer, J M; Prakash, M; Steiner, A W. Neutrino Emission from Cooper Pairs and Minimal Cooling of Neutron Stars. {\it Astrophys. J.} {\bf 2009}, 707, 1131.

\bibitem{urca} Lattimer, J M; Pethick, C J; Prakash, M; Haensel, P. Direct URCA Process in Neutron Stars. {\it Phys. Rev. Lett.} {\bf 1991}, 66, 2701.
\bibitem{urca2} Thapa, V B; Sinah, M. Direct URCA process in light of PREX-2.  arXiv:2203.02272 [nucl-th].

\bibitem{Malik22} Malik, T; Agrawal, B K; Providencia C. Inferring the nuclear symmetry energy at supra saturation density from neutrino cooling.  arXiv:2206.15404v2  [nucl-th].

\bibitem{Malik22_2} Patra, N K; Iman, Sk Md A; Agrawal, B K;  Mukherjee, A; Malik, T. Nearly model-independent constraints on dense matter equation of state. {\it Phys. Rev. D} {\bf 2022}, 106, 043024.
in a Bayesian approach

\bibitem{FS_2022} Sammarruca, F. Neutron skin systematics from microscopic equations of state. {\it Phys. Rev. C} {\bf 2022}, 105, 064303.



\bibitem{DHW21} Drischler, C; Holt, J W; Wellenhofer, C. Chiral Effective Field Theory and the High-Density Nuclear Equation of State. {\it Annu. Rev. Nucl. Part. Sci.} {\bf 2021}, 71, 403. 



\bibitem{LH19} Lim, Y;  Holt, J. Bayesian modeling of the nuclear equation of state for neutron star tidal deformabilities and GW170817.  {\it Eur. Phys. J. A} {\bf 2019}, 55, 209. 

\bibitem{crex} Reinhard, P-G; Roca-Maza, X; Nazarewicz, W. Combined Theoretical Analysis of the Parity-Violating Asymmetry for $^{48}$Ca and $^{208}$Pb. {\it Phys. Lett.} {\bf 2022}, 129, 232501. 
\bibitem{msu}
https://frib.msu.edu/news/2022/prl-paper.html  


\end{references}
\end{document}